\documentclass[aps,prl,twocolumn,amsfonts,showpacs,longbibliography]{revtex4-2}
\usepackage{verbatim,epsfig,amsmath,amssymb,bm,epsf,graphicx,psfrag,bbold,amsthm,amsfonts}
\usepackage[bottom]{footmisc}
\usepackage{hyperref}
\usepackage{cleveref} 
\usepackage{enumitem}
\usepackage{grffile}
\usepackage{framed}
\usepackage{mathrsfs}
\usepackage{esint}
\setlist{nosep}
\usepackage{placeins}
\usepackage[all]{xy}
\usepackage{color}
\usepackage[utf8]{inputenc}
\usepackage{float}
\usepackage{natbib}
\usepackage{tikz-cd}
\usepackage{verbatim}
\usepackage{leftidx}
\usepackage[normalem]{ulem}

\usepackage{notes2bib}

\newcommand{\moy}[1]{\langle #1 \rangle}
\newcommand{\ket}[1]{\mbox{$ | #1 \rangle $}}

\newcommand\bZ {{\mathbb Z}}

\newcommand\beq {\begin{equation}}
	\newcommand\eeq {\end{equation}}
\newcommand\beqa {\begin{equatiobn}\begin{array}}
		\newcommand\eeqa {\end{array}\end{equation}}
\newcommand\bal {\begin{align}}
	\newcommand\eal {\end{align}}
\newcommand{\bea}{\begin{eqnarray}}
	\newcommand{\eea}{\end{eqnarray}}

\newcommand{\ztwo}{\mathbb{Z}_2}

\theoremstyle{plain}

\theoremstyle{definition}

\theoremstyle{remark}

\usepackage{pdfpages} 
\usepackage{pgffor} 

\makeatletter
\AtBeginDocument{\let\LS@rot\@undefined}
\makeatother

\def\supplementfilename{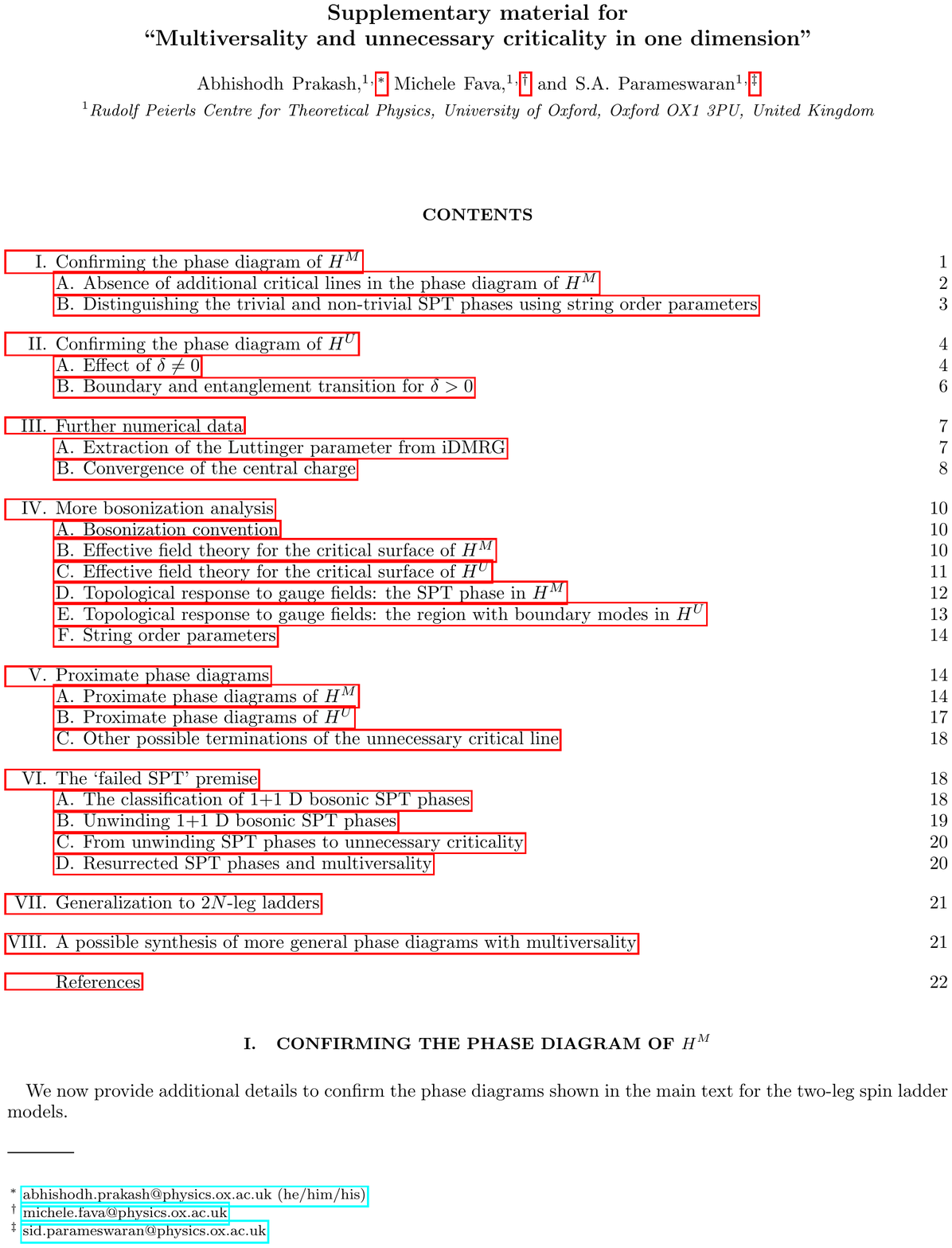}

\pdfximage{\supplementfilename}
\def\numbersupplementpages{\the\pdflastximagepages}

\newif\ifarXiv
\arXivtrue 

\begin{document}
	
	\title{Multiversality and Unnecessary  Criticality in One Dimension}
	\author{Abhishodh Prakash}
	\email{abhishodh.prakash@physics.ox.ac.uk (he/him/his)}
	\affiliation{Rudolf Peierls Centre for Theoretical Physics, University of Oxford, Oxford OX1 3PU, United Kingdom}
	
	\author{Michele Fava}
	\email{michele.fava@physics.ox.ac.uk}
	\affiliation{Rudolf Peierls Centre for Theoretical Physics, University of Oxford, Oxford OX1 3PU, United Kingdom}
	
	\author{S.A. Parameswaran}
	\email{sid.parameswaran@physics.ox.ac.uk}
	\affiliation{Rudolf Peierls Centre for Theoretical Physics, University of Oxford, Oxford OX1 3PU, United Kingdom}
	

	\begin{abstract}
		We present microscopic models of spin ladders which exhibit continuous critical surfaces whose  properties and existence, unusually, cannot  be inferred from those of the flanking phases. These models exhibit either ``multiversality" --  the presence of different universality classes over finite regions of a critical surface separating two distinct phases -- or its close cousin, ``unnecessary criticality"-- the presence of a stable critical surface within a single, possibly trivial, phase. We elucidate these properties using Abelian bosonization and density-matrix renormalization-group simulations, and attempt to distill the key ingredients required to generalize these considerations.
	\end{abstract}
	
	\maketitle

	Quantum criticality~\cite{Sachdev_book,SondhiReview} plays a central role in our understanding of  zero-temperature phases 
	of matter. The existence of critical points or surfaces can usually be inferred even without probing the transition region, by observing suitably distinct quantum ground states in disjoint parameter regimes.  When continuous, their universal scaling  properties  are likewise assumed to be uniquely determined by the flanking phases, unless fine-tuned. These ideas are thought to hold even when the Landau picture of broken symmetries is modified to include topological distinctions between phases, or in transitions, such as those proposed between  distinct broken-symmetry orders, whose  fluctuating critical degrees of freedom are not natural excitations of either adjacent phase~\cite{Senthil_DQC_2004}. 
	\begin{figure}[t!]
		\centering
		\includegraphics[width=0.48\textwidth]{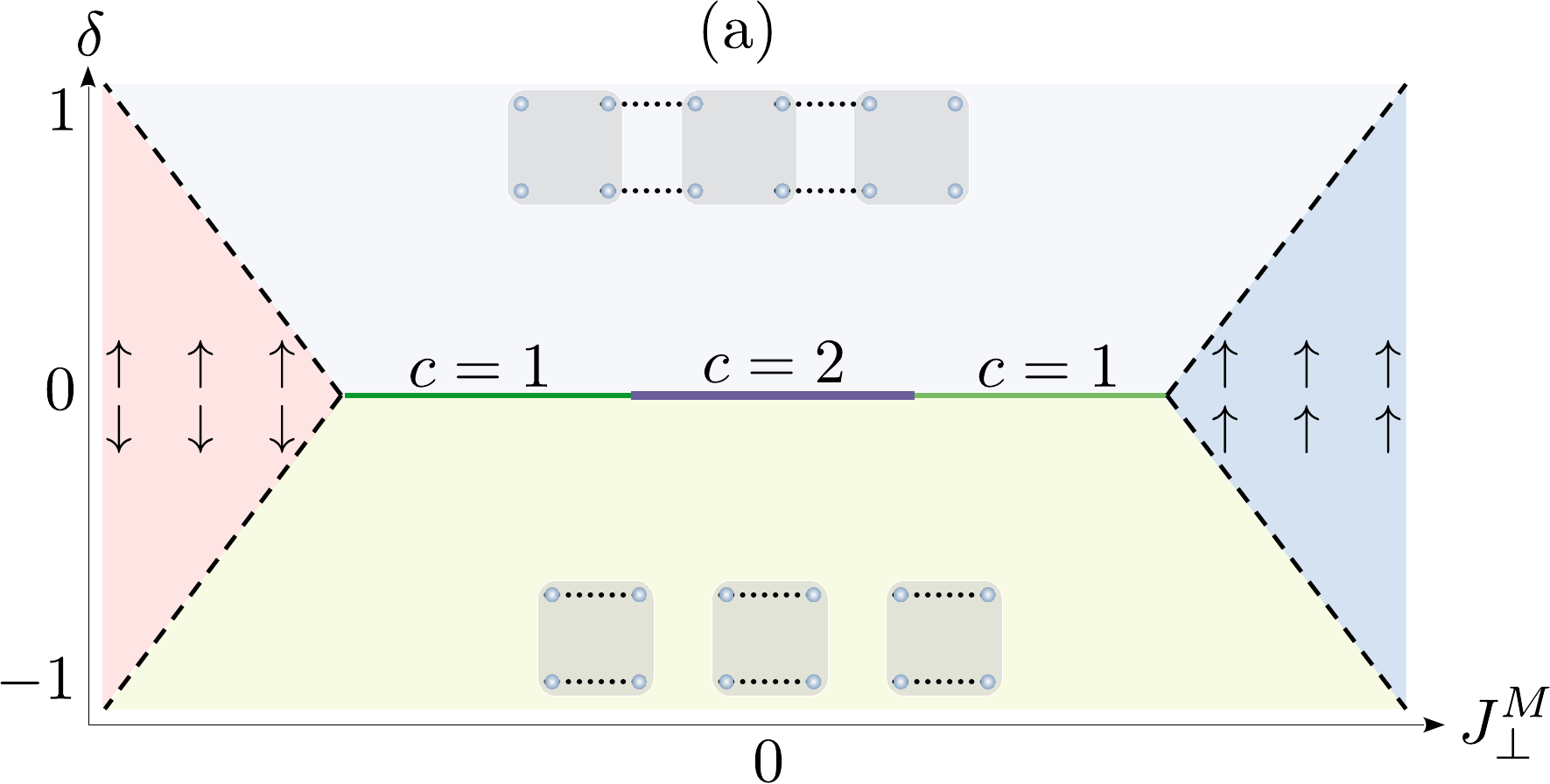}
		\includegraphics[width=0.48\textwidth]{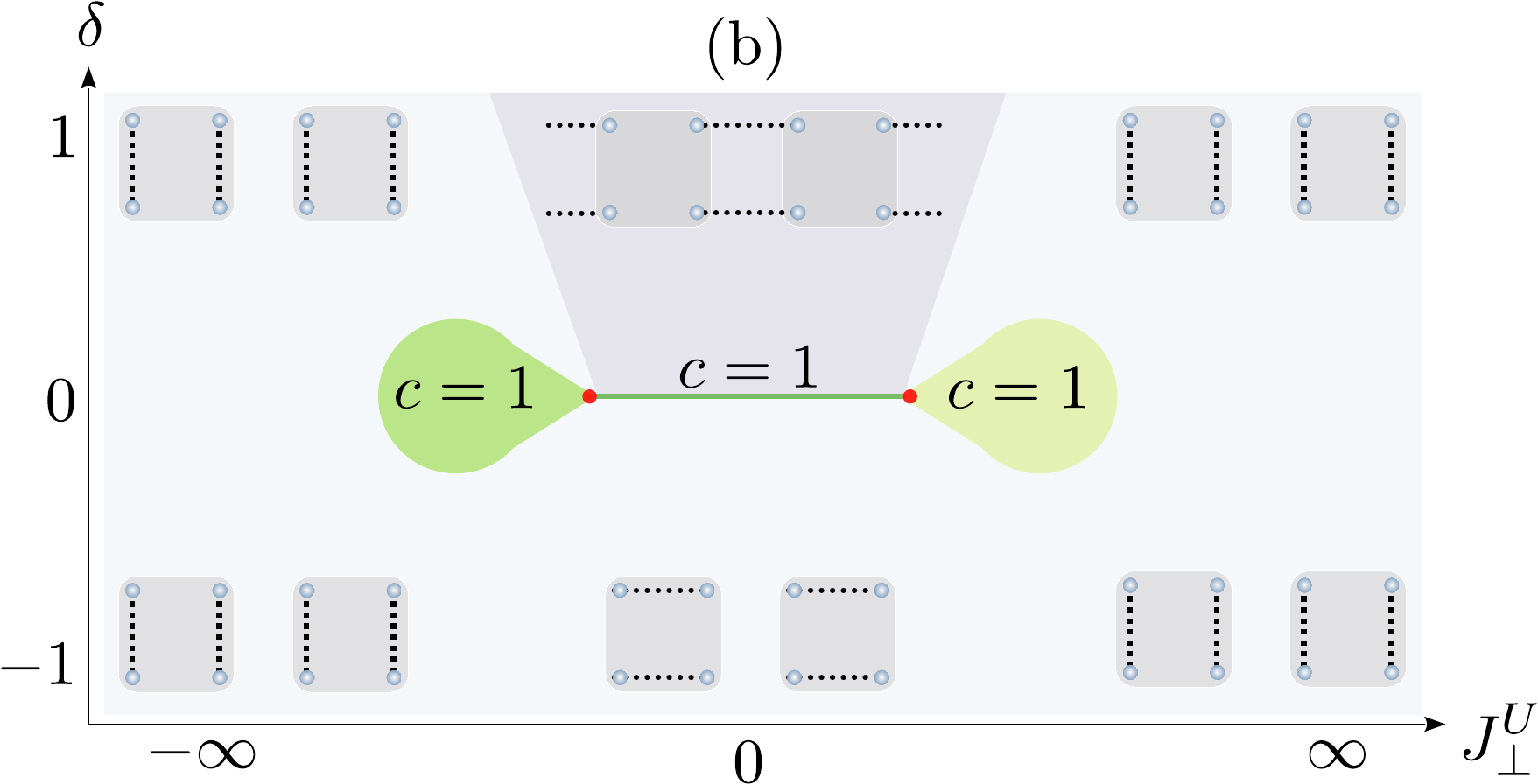} 
		\caption{\textbf{(a)}: The Hamiltonian $H^M$, obtained by perturbing  $H_\delta$ by $H^M_\perp$, [cf \cref{eq:Hdelt,eq:HperpN1_multi}, $\Delta\in \left(-\frac{1}{\sqrt{2}},0\right)$], yields a  phase diagram exhibiting multiversality. Solid lines along the $\delta$ = 0 line denote distinct universality classes  separating the same two gapped phases, as reflected by the different central charges $c$ of the respective conformal field theories. Dashed lines denote first-order phase transitions. \textbf{(b)}: If instead  $H_\delta$ is perturbed by $H^U_\perp$ [cf \cref{eq:HperpN1_uc}, with $J^M_\perp <0$ fixed to a value leading to the $c=1$ segment in \cref{fig:Phasediags}.a], the phase diagram hosts an unnecessary critical line along $\delta = 0$ with  $c=1$. The critical line terminates into into lobes of distinct $c=1$ critical Luttinger liquid phases through a $c= \frac{3}{2}$ phase transition (red dot). The shaded region hosts stable boundary modes and is separated from the rest of the phase diagram by a boundary transition.}
		\label{fig:Phasediags}
	\end{figure}
	
	Recently, attention has focused on a pair of converse questions: namely (1) whether a phase distinction is necessary for a critical surface to exist; and (2) when two distinct phases straddle a critical surface, if this distinction uniquely fixes the universality class of the  transition between them. Surprisingly, the answer to both these questions is in the negative. First, it is possible to have   a critical surface within the \textit{same} phase, accessed by tuning a single parameter, whose presence is  not demanded by phase structure. Second, there exist  generic (i.e. non-fine-tuned) transitions between the same pair of phases with distinct universality classes depending on the path in parameter space taken across the critical surface. These phenomena have been identified in a handful of models, usually invoking topology in an essential way. The first class of models with ``unnecessary  criticality''~\cite{AnfusoRosch_PhysRevB.75.144420,MoudgalayaPollman_PhysRevB.91.155128,BiSenthil_PhysRevX.9.021034,Pollmann_Quotient} can emerge upon modifying symmetries to remove a topological distinction~\cite{WangWenWitten_Unwinding_PhysRevX.8.031048,AP_UnwindingBosonic_PhysRevB.98.125108,AP_UnwindingSUSY_PhysRevB.103.085130,AP_SUSY_PhysRevLett.126.236802} between two phases: a continuous critical surface required by the distinction becomes unnecessary in its absence. Identifying criteria for the second phenomenon --- evocatively dubbed ``multiversality'' by Bi and Senthil~\cite{BiSenthil_PhysRevX.9.021034} ---  is more elusive. An early classical example leveraged topological distinctions within the disordered phase of an XY model in $d=2$ spatial dimensions augmented with half-vortex defects~\cite{Fendley_KTIsing_PhysRevLett.107.240601,SernaChalkerFendley_KTIsing_2017}. More recent  quantum settings involve  Dirac fermions perturbed by topological mass terms {and strong interactions} in  $d=2$~\cite{SlagleYouXu_PhysRevB.91.115121} or coupled to fluctuating non-Abelian gauge fields in $d=3$~\cite{BiSenthil_PhysRevX.9.021034}. However, the critical field theories and scaling properties of these  examples of multiversality and unnecessary criticality can be  challenging to access analytically or even numerically, especially in the $d>1$ quantum setting. It is thus desirable to identify  microscopic models that exhibit both  phenomena in an analytically tractable regime, ideally in $d=1$ where the density-matrix renormalization group (DMRG) allows accurate numerical simulations.

	Here, we show that both phenomena arise in $d=1$ spin ladder models, 
	that can be accessed  analytically via (Abelian) bosonization, and numerically via DMRG. We employ both strategies to map out their phase structure, and comment both on their relation to existing work and the possibility of generalizing these ideas to a broader set of models. Our work thus provides a  basis for deeper investigations of the link between symmetry-protected topological order, criticality, and phase structure, and suggests the ingredients needed to identify further instances of multiversality and unnecessary criticality.
	
	\medskip

	\noindent\emph{Models.---} We begin with a class of two-leg ladder  Hamiltonians of the form  
	$H^{M/U} = H_\delta+  H_\perp^{M/U}$. Here,
	\begin{align}
		H_\delta &= \sum_{j=1}^L\sum_{\alpha=1,2} (1+\delta(-1)^j) h_{\alpha j}, \text{ with }
		\nonumber\\
		h_{\alpha j} &= S^x_{\alpha j} S^x_{\alpha j+1}  +  S^y_{\alpha j} S^y_{\alpha j+1} + \Delta  S^z_{\alpha j} S^z_{\alpha j+1},\label{eq:Hdelt}
	\end{align}
	where $\vec{S}_{\alpha j} = \frac{1}{2} \vec{\sigma}_{\alpha j}$  are spin-$\frac12$ operators written in terms of Pauli matrices ${\sigma}^\mu_{\alpha j}$. $H_\delta$ describes two identical decoupled XXZ spin chains, whose couplings are staggered when $\delta\neq 0$.
	We  fix the anisotropy $\Delta\in \left(-\frac{1}{\sqrt{2}},0\right)$, for reasons discussed below. 
	In this decoupled limit, each leg hosts two gapped phases: a trivial paramagnet for $\delta<0$ and a symmetry-protected topological (SPT) phase~\cite{Senthil_SPTReview_2015,ChiuSPTReview__2016} with gapless  boundary modes -- related to the celebrated Haldane phase~\cite{Haldane_NLSM_PhysRevLett.50.1153,AKLT_PhysRevLett.59.799,KohmotoNijsKadanoff_XXZDimerization_PhysRevB.24.5229} -- for $\delta>0$, separated by a continuous transition at $\delta=0$. In the fully dimerized, fully decoupled limits $\delta = \pm 1$  the exact ground states of $H_\delta$ are 
	\begin{align}
		\ket{\text{GS}(\delta=\pm1, J_\perp=0)} = \prod_{\alpha=1,2} \prod_{j \in \mathcal{J}_\pm}\ket{[\alpha, j; \alpha, j+1]}, \label{eq:GS_delta}
	\end{align}
	where $\ket{[\alpha,i;\beta,j]}$ represents an SU(2) singlet entangling sites $(\alpha,i)$ and $(\beta,j)$, and $\mathcal{J}_+$ and $\mathcal{J}_-$ denote the set of even and odd sites, respectively.  On Abelian bosonization of $H_\delta$~\cite{Giamarchi}, keeping only the most relevant terms, we have 
	\begin{multline}
		H_\delta  \approx  \frac{v}{2 \pi} \int dx \sum_{\alpha = 1}^{2} \left[\frac{1}{4K} \left(\partial_x \phi_\alpha\right)^2 + K \left(\partial_x \theta_\alpha\right)^2\right]\\ +  \mathcal{A}^2 \delta  \int dx~    \left(\cos \phi_1 + \cos \phi_2\right) \label{eq:Hdeltabosonized}
	\end{multline}
	where $\phi_\alpha \cong \phi_\alpha + 2 \pi $ and $\theta_\alpha \cong \theta_\alpha + 2 \pi $ are canonically conjugate {compact} boson fields satisfying $[\partial_x \phi_\alpha(x),\theta_\beta(y)] = 2 \pi i \delta_{\alpha \beta} \delta(x-y)$, $\mathcal{A}$ is a bosonization prefactor whose precise value is unimportant, {{and}}
	the Luttinger parameter $K = \frac{\pi}{2 }\left(\pi-\arccos \Delta\right)^{-1}$ and velocity $v= \frac{K}{(2K-1)} \sin \left(\frac{\pi}{2K}\right)$ are determined from the Bethe ansatz solution of the XXZ chain~\cite{Haldane_BetheLL_1981153}. For $\Delta \in \left(-\frac{1}{\sqrt{2}},0\right)$, we have $K \in \left(1,2\right)$ and thus the vertex operators $\mathcal{U}_{1,2} \equiv \cos\phi_{1,2}$, which have scaling dimensions
	\begin{equation}
		\left[\mathcal{U}_1\right] = \left[\mathcal{U}_2\right] = K,
	\end{equation}
	are relevant~\footnote{Recall that an operator in a conformal field theory (CFT) is relevant if its scaling dimension is lower than the space-time dimension} and open a gap for any $\delta \neq 0$, pinning the fields at $\moy{\phi_{1,2}} = \frac{\pi}{2} (1+ \text{sgn}(\delta))$. Thus, the bosonized description recovers the  $J_\perp=0$ phase structure discussed above, with a critical point at $\delta=0$. The decoupled  model enjoys an O(2)$\times$ O(2) symmetry generated by independent $U(1)$ spin rotations $S^{\pm}_{\alpha j} \mapsto e^{\pm i \chi_\alpha}S^{\pm}_{\alpha j}$ and spin reflections $ \{S^{\pm}_{\alpha j} \mapsto S^{\mp}_{\alpha j},~S^{z}_{\alpha j} \mapsto - S^{z}_{\alpha j} \}$  on each leg, and $\ztwo$ leg exchange symmetry  $\vec{S}_{1j} \leftrightarrow \vec{S}_{2j}$ which enforces the critical points for both legs to coincide.

	We now show analytically and verify numerically that  introducing two distinct forms of interlayer coupling that preserve different subsets of these symmetries,
	\begin{align}
		H_\perp^M &=  J^M_\perp \sum_j  S^z_{1 j} S^z_{2 j}, \label{eq:HperpN1_multi} \\
		H_\perp^U &=   J^M_\perp \sum_j  S^z_{1 j} S^z_{2 j} + J^U_\perp \sum_j \left(S^x_{1 j} S^x_{2 j}+S^y_{1 j} S^y_{2 j}  \right), \label{eq:HperpN1_uc}	
	\end{align}	
	leads to the phase diagrams in Fig.~\ref{fig:Phasediags} that respectively exhibit multiversality and unnecessary criticality.
	
\begin{figure}[!t]
		\centering
		\includegraphics[width=0.48\textwidth]{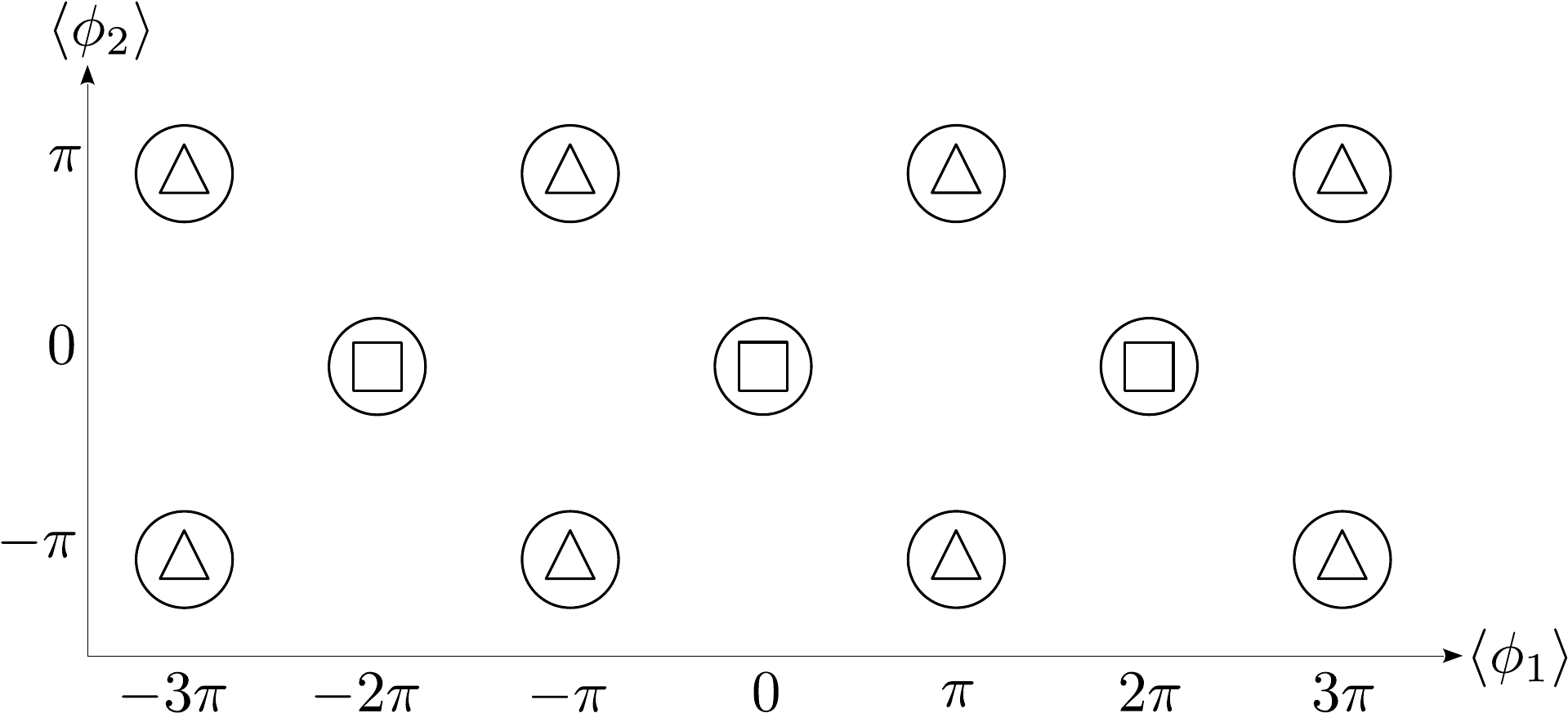} 
		\caption{Values of $\moy{\phi_{1,2}}$ pinned by $H_\delta$ for $\delta <0$ (squares), $\delta >0$ (triangles)  and by $H_\perp^M$ (circles). Since the triangles and squares are a subset of the circles, $H_\delta$ and $H^M_\perp$ have compatible ground states.}
		\label{fig:vacua}
	\end{figure}

	\medskip 
	\noindent \emph{Multiversality.---} $H_\perp^M$ preserves layer exchange and independent spin rotations but only retains simultaneous spin reflections thereby breaking the on-site O(2)$\times$ O(2) symmetry down to (U(1)$\times$ U(1)) $\rtimes \ztwo$.  This preserves the $J^M_\perp=0$ phase structure although it reduces the degeneracy of boundary modes in the non-trivial SPT phase, as the system crosses over from a O(2)$\times$ O(2)  SPT phase to a (U(1)$\times$ U(1)) $\rtimes \ztwo$  SPT phase without any bulk phase transition. To study the effect of $H_\perp^M$ on the $\delta=0$ critical point, we consider its  bosonized form ($\mathcal{B}$ is another bosonization prefactor),
	\begin{multline}
		H_\perp^M \approx  \mathcal{B}^2 J^M_\perp \int dx\left(  \cos \left(\phi_1 - \phi_2\right) -   \cos \left(\phi_1 + \phi_2\right) \right) \\+ \frac{J^M_\perp}{4 \pi^2}  \int dx \left( \partial_x \phi_1 \partial_x \phi_2 \right)   ,   \label{eq:HperpN1_multi_bosonized}
	\end{multline}
	from which we see that it introduces two new vertex operators, $\mathcal{V}_{\pm} \equiv \cos \left( \phi_1 \pm \phi_2\right)$ which   involve combinations of the boson fields that are respectively symmetric and antisymmetric under layer exchange. For $J^M_\perp=0$, both $\mathcal{V}_\pm$  have scaling dimension $2K$, and are hence irrelevant for our choice of  $\Delta \in \left(-\frac{1}{\sqrt{2}},0\right)$: the critical theory remains a $c=2$ two-component Luttinger liquid  for small  $|J^M_\perp|$. As $|J^M_\perp|$ is increased, it changes operator scaling dimensions through its coupling to the exactly marginal operators $\partial_x\phi_1 \partial_x\phi_2$:  perturbatively in $J^M_\perp$, 
	\begin{equation}
		[\mathcal{V}_\pm]\equiv K_\pm \approx 2K \left(1 \mp \frac{J^M_\perp K}{2 \pi v}\right), \label{eq:Klut_pert}
	\end{equation}
	suggesting that $\mp\cos \left( \phi_1 \pm \phi_2\right)$  become relevant for the critical values $\pm J^{M*}_\perp \approx \pm \frac{2 \pi v (K-1)}{K^2} $ respectively and gap out either the leg-symmetric or leg-antisymmetric components of the Luttinger liquid. The resulting single-component Luttinger liquid   corresponds to a critical theory with $c=1$. In order to show that we have the multiversal line shown in \cref{fig:Phasediags}a, we must verify that the $\delta \neq 0$ gapped phases remain unchanged away from the critical line as we turn on $J^M_\perp$ . To do so, we first observe that although the scaling dimensions of 
	$\mathcal{U}_{1,2}$ are modified to
	\begin{equation}
		\left[\mathcal{U}_{1,2}\right] = \begin{cases}
			~~~\frac{K_+}{4} &\text{ for } J^M_\perp < -J^{M*}_\perp\\	
			\frac{K_+ + K_-}{4} &\text{ for }|J^M_\perp| < |J^{M*}_\perp|\\
			~~~\frac{K_-}{4}&\text{ for }J^M_\perp > +J^{M*}_\perp
		\end{cases} , \label{eq:Hdelta_Klut} 
	\end{equation}
	from \cref{eq:Klut_pert}, they remain relevant as $J^{M}_\perp$ is tuned through  $\pm J^{M*}_\perp$. Now,  $H_\delta$ pins the fields to the values $\moy{\phi_{1,2}} = \frac{\pi}{2} (1+ \text{sgn}(\delta))$ (the minima of $\mathcal{U}_{1,2}$), while $H_\perp^M$ pins $\moy{\phi_1 + \text{sgn}(J_\perp^M) \phi_2} = 2 \pi \bZ$ for $|J_\perp^M| > |J^{M^*}_\perp|$ (the minima of $\mathcal{V}_{\pm}$). Since the minima of $H_\perp^M$ are compatible with those of $H_\delta$ (see \cref{fig:vacua}), 
	there will be no qualitative change in the nature of the $\delta \neq 0$ gapped phases across $J^{M^*}_\perp$.  
	
	The nature of the ordered phases at large $|J^M_\perp|$ are easily determined from first-order perturbation theory on $H_M$~\cite{supp}. Since the ground states of these ordered phases belong to a different total layer-magnetization sector $S^z_{\text{tot,}\alpha} = \sum_j S^z_{\alpha j}$, 
	we expect a first-order transition between them and the original small-$J_\perp^M$ gapped phases.
	
	Combining these results, we obtain the phase diagram in Fig.~\ref{fig:Phasediags}a with a multiversal critical  line,  described  by a $c=2$ or $c=1$ conformal field theory (CFT) depending on the path taken in $(\delta, J^M_\perp)$ space between the two straddling phases. 
	
\begin{figure}[!t]
		\begin{tabular}{c}
			\includegraphics[width=0.48\textwidth]{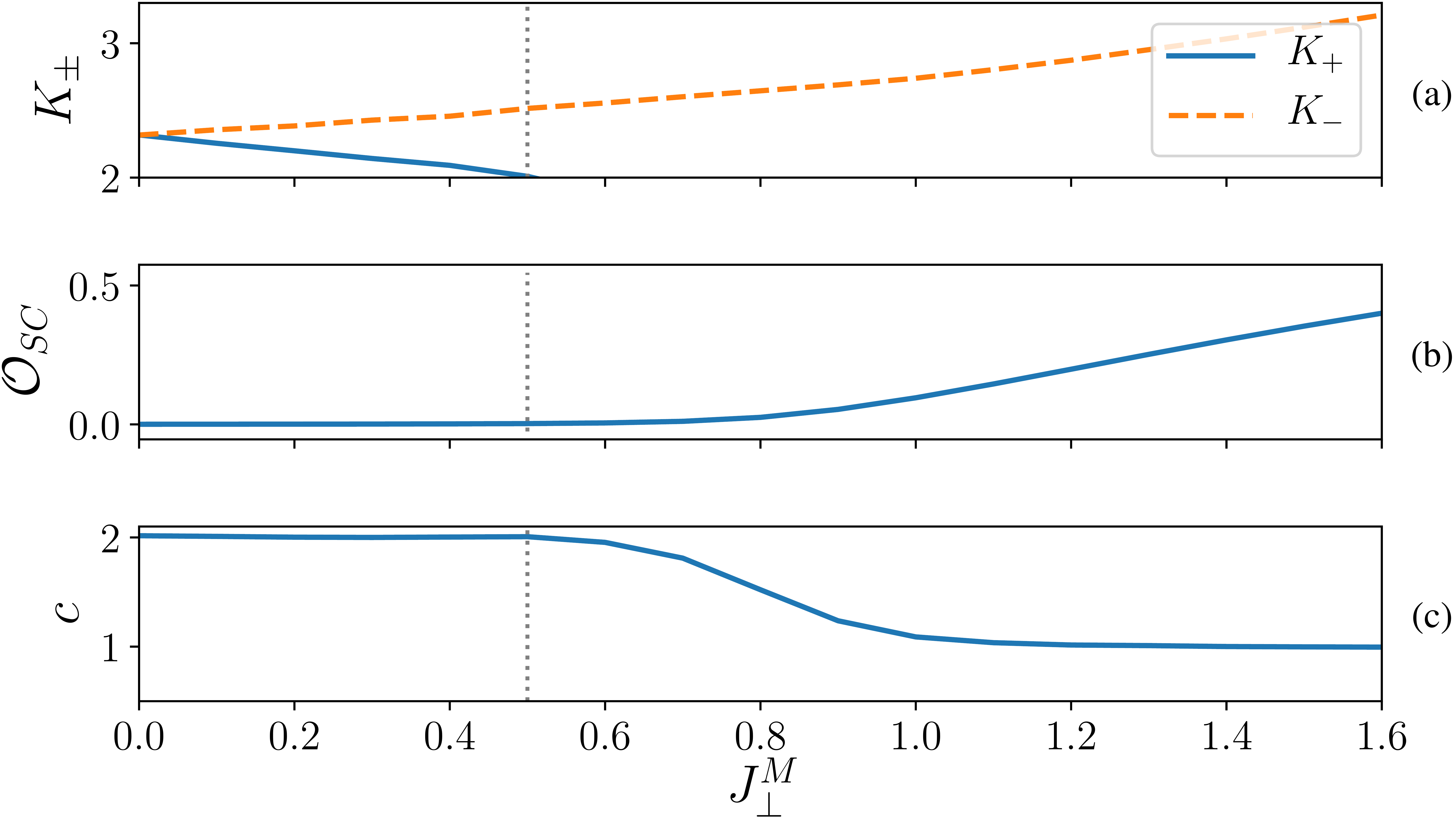} 
		\end{tabular}
		\caption{Multiversality. From top to bottom-- (a) Luttinger parameters $K_\pm$, (b) string order parameter $\mathcal{O}_{\text{SC}}$ and  (c) central charge $c$  computed along the $\delta=0$ line of $H^M$ for various $J^M_\perp>0$ with fixed $\Delta = -0.25$ using iDMRG~\cite{supp}. The vertical dotted line ($J^{M*}_\perp \approx 0.5$) denotes where $K_+ =2$ and the critical CFT changes from $c=2$ to $c=1$.}
		\label{fig:SPT1}
	\end{figure}

	This phase diagram can be numerically verified via IDMRG~\cite{White_DMRG_PhysRevLett.69.2863}, most efficiently by restricting  attention to the $\delta=0$ line~\cite{supp} and $J^M_\perp>0$ (the latter since $J^M_\perp \rightarrow -J^M_\perp$ is a unitary transformation). We extract $K_\pm$, the scaling dimensions of $\mathcal{V}_\pm$, via the correlation functions of two independent scaling operators $S^+_{1j} S^\pm_{2j} \sim e^{i (\theta_1 \pm \theta_2 )}$~\cite{supp} and the central charge $c$ through finite-entanglement scaling~\cite{PollmannMukherjeeTurnerMoore_CFT_S_PhysRevLett.102.255701,Verstraete_EntanglementScaling_PhysRevB.91.035120}. We also use the  string order parameter
	\begin{equation}
		\mathcal{O}_{\text{SC}} = \lim_{r \rightarrow \infty}  \biggr<\prod_{l=j}^{j+r}  \sigma^z_{1l}  \sigma^z_{2l} \biggr>, \label{eq:String_orderpar}
	\end{equation}
	which picks up an expectation value when $\moy{\phi_1 \pm \phi_2} = 0$~\cite{NAKAMURA_string,supp} and the central charge changes to $c=1$. By tracking the evolution of $K_\pm$, $ \mathcal{O}_{\text{SC}}$ and $c$ along the $\delta = 0$ line (\cref{fig:SPT1}), we see that the central charge drops from  $c=2$ to $c=1$ as  $J_\perp^M$ is tuned through  $\pm J^{M^*}_\perp$ while  
	$H_\delta$ remains relevant [cf \cref{eq:Hdelta_Klut}]. 
	
	\medskip 
	
	\noindent \emph{Unnecessary criticality.---} We now consider $H_\perp^U$ in \cref{eq:HperpN1_uc}, with a fixed value of $J^M_\perp < -J^{M^*}_\perp$ (such that $c=1$ when $J_\perp^U=0$, cf \cref{fig:Phasediags}a). This preserves the leg-exchange symmetry of $H^M_\perp$ but breaks the on-site symmetry down to the O(2) generated by simultaneous spin rotations 
	and reflections in both legs. $H_\perp^U$ eliminates the distinction between the gapped regions of $H^M_\perp$~\cite{supp} for different signs of $\delta$. The easiest way to see this is by observing that both exact ground states in \cref{eq:GS_delta}  evolve to the \textit{same} product state as  $J^U_\perp \rightarrow \infty$, without a bulk phase transition:
	\begin{equation}
		\ket{GS(\delta = \pm 1, J^U_\perp \rightarrow \infty )} = \prod_{j=1}^L \ket{[1,j;2,j]},
	\end{equation} 
	where $\ket{[1,j;2,j]}$ denotes a singlet along the j$^\text{th}$ rungs of the ladder (\cref{fig:Phasediags}b). A similar result obtains 
	for $J_\perp^U \rightarrow - \infty$ but with the singlet replaced by a different entangled Bell pair. 
	This implies that there is a single gapped phase in the periphery of the entire $(\delta,J^U_\perp)$ region. To determine the fate of the system closer to the origin $\delta=J^U_\perp=0$, we use the bosonized version of $H_\perp^U$, 
	\begin{equation}
		H_\perp^U \approx H_\perp^M + J^U_\perp \mathcal{C}^2\int dx ~   \cos(\theta_1 - \theta_{2}), \label{eq:H_bosonized_UC}
	\end{equation}
	with $H_\perp^M$ as in \eqref{eq:HperpN1_multi_bosonized} and $\mathcal{C}$ again an unimportant prefactor. 
	\begin{figure}[!t]
		\centering
		\includegraphics[width=0.48\textwidth]{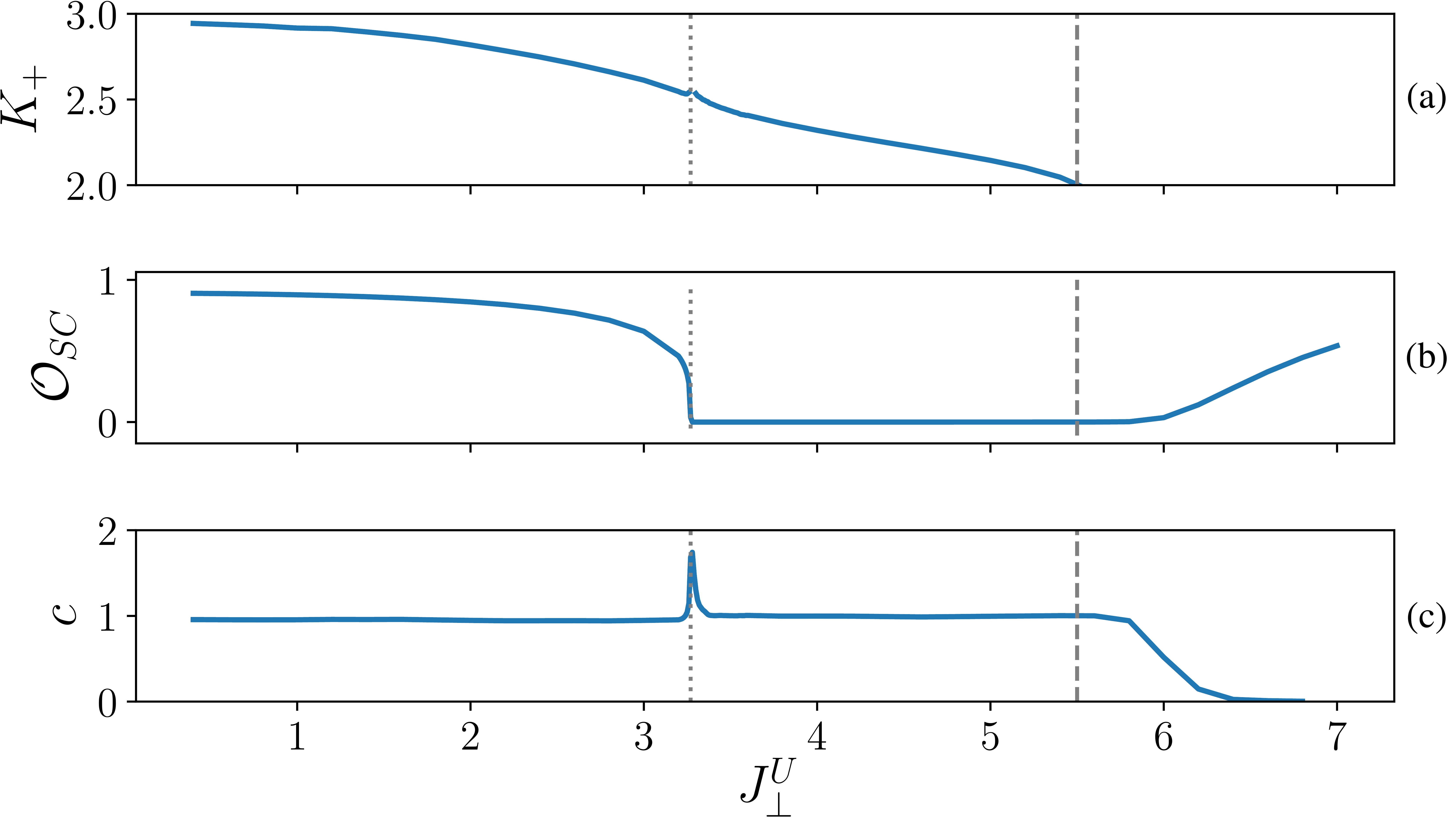} 
		\caption{Unnecessary Criticality. From top to bottom-- (a) Luttinger parameter $K_+$, (b) string order parameter $\mathcal{O}_{\text{SC}}$ and  (c) central charge $c$  along the $\delta=0$ line of $H^U$  for various $J_\perp^U>0$ with fixed $\Delta = -0.05$ and $J^M_\perp = -5.2$ using iDMRG~\cite{supp}. The dotted  line ($J^U_\perp \approx 3.27$) denotes the $c=\frac{3}{2}$ point when the XY$_2$ critical line transitions to XY$_1$ lobes . The dashed line ($J^U_\perp \approx 5.5$) denotes the point where $K_+=2$ when the system transitions to a trivial gapped phase.}
		\label{fig:UC1}
	\end{figure}
	For nonzero $J_\perp^M$ and $J_\perp^U$, 
	at least one of  $\mathcal{W}_- \equiv \cos(\theta_1 - \theta_2)$ or $\mathcal{V}_-$ 
	is always relevant; if $\mathcal{V}_+$ 
	is irrelevant the system flows to a gapless $c=1$ theory. However, the nature of this theory depends on which of the two operators dominates at large distances.
	When $\mathcal{W}_-$ 
	dominates, ${\theta_1 - \theta_2}$ is pinned, while ${\phi_1-\phi_2}$ fluctuates. Instead, when $\mathcal{V}_-$
	dominates, ${\theta_1 - \theta_2}$ fluctuates while ${\phi_1-\phi_2}$ is pinned.  Following the terminology of Ref.\cite{Schulz_Spin1_PhysRevB.34.6372}, we  refer to these Luttinger liquids as XY$_1$ and XY$_2$ respectively.
	We find that there exists a range of values of fixed $\Delta \in \left(-1,0\right)$ and $J^M_\perp<-J^{M*}_\perp$ such that we get a stable extended XY$_2$ unnecessary critical line extending from the origin along $\delta = 0$. For $\delta \neq 0$ away from this line, $\mathcal{U}_{1,2}$ are relevant and drive the system to a gapped phase. While there are other possible ways for this line to terminate~\cite{supp}, for the chosen parameters, the XY$_2$ line first transitions to XY$_1$ on each end and then terminates. In the XY$_1$ regions, $\mathcal{U}_{1,2}$ decay exponentially and $H_\delta$ cannot gap out the system. As a result, the XY$_1$ line opens up into small islands of gapless \textit{phases}, that persist until $\mathcal{V}_+$ becomes relevant and drives the system to a fully gapped  trivial phase. The two  XY$_1$  lobes on each end of the critical line correspond to $\moy{\theta_1-\theta_2} = 0$ or  $\pi$ (depending on the sign of $J_\perp^U$) in fact represent distinct phases which cannot be connected without a phase transition due to different symmetry charges carried by the gapless degrees of freedom~\cite{supp,toappearAP}. 
	
	A schematic of the phase diagram is shown in \cref{fig:Phasediags}b. We can once again numerically verify all aspects of the phase diagram via DMRG at $\delta=0$ 
	(Fig.~\ref{fig:UC1}). $\mathcal{O}_{\text{SC}}$  (the same string operator defined in \cref{eq:String_orderpar})  now picks up an expectation value in the XY$_2$ critical region and in the trivial gapped phase, but not the XY$_1$ lobes. Since $J^U_\perp \mapsto -J^U_\perp$ is a unitary transformation, we restrict our attention to $J^U_\perp>0$.
	We see that indeed a stable XY$_2$ line with $c = 1$ persists until it transitions to XY$_1$. In the numerics this is marked by a jump in the central charge: indeed the transition from XY$_1$ to XY$_2$ is known to happen through a $c = \frac{3}{2}$ critical point corresponding to the gapped sector undergoing an Ising transition~\cite{LECHEMINANT_SelfDualSineGordon_2002502}. Finally, at larger values of $J_\perp^U$, $K_+$ dips below 2 and the system gaps out.
	
	We have thus  embedded a $c=1$ CFT as an unnecessary critical line  not demanded by the phase structure (as there is a unique gapped phase in the phase diagram) that can be accessed by tuning a single parameter.		
	
	\medskip 
	\noindent\emph{Boundary transitions meet the bulk}: A curious feature of the phase diagram of $H^U_\perp$ is the presence of stable boundary modes above the unnecessary critical line. This can be seen in the limiting case  $\delta = 1$, where the effective boundary Hamiltonian on each end acts on two spins and has the form (suppressing site labels for brevity)
	\begin{equation}
		H_{\partial} = J^M_\perp  S^z_{1} S^z_{2} + J^U_\perp \left(S^x_{1 } S^x_{2 }+S^y_{1 } S^y_{2}  \right).
	\end{equation}
	This has a twofold degenerate ground state for $|J^U_\perp|<-J^M_\perp$ so that $H^U$ has boundary modes, and a unique ground state for $|J^U_\perp|>-J^M_\perp$ so that $H^U$ has no boundary modes. We can numerically verify~\cite{supp} that the boundary modes are stable even as we reduce $\delta$ (shaded region in \cref{fig:Phasediags}b). Remarkably, the boundary transition (at $|J^U_\perp|= -J^M_\perp$ for $\delta = 1$) terminates at the $c=\frac{3}{2}$ point, the same as the unnecessary critical line. If the bulk and boundary transitions are treated on equal footing, the unnecessary critical line becomes part of a phase boundary separating ``boundary-obstructed" topological phases~\cite{Khalaf_BoundaryObstructed_PhysRevResearch.3.013239}, leading to a more  conventional-looking phase diagram. 
	
	We conjecture that unnecessary critical lines in the bulk generically terminate by turning into boundary critical lines and enclose regions with stable boundary modes. While this is true in all known one-dimensional examples~\cite{AnfusoRosch_PhysRevB.75.144420,Pollmann_Quotient}, it would be interesting to verify in higher-dimensional examples too~\cite{BiSenthil_PhysRevX.9.021034,JianXu_PhysRevB.101.035118}, as it suggests an intriguing universal connection between unnecessary criticality, boundary criticality, and stable gapless modes. 
	
	\medskip 
	\noindent \emph{Stability of phase diagrams:} The field theories shown in \cref{eq:Hdeltabosonized,eq:HperpN1_multi_bosonized,eq:H_bosonized_UC} already contain the most relevant symmetry-allowed scaling operators. As a result, the phase diagrams shown in \cref{fig:Phasediags} are stable to arbitrary (but small) symmetry allowed perturbations and small variations of existing parameters. These can only introduce corrections to parameters of the field theory which in turn only quantitatively change \cref{fig:Phasediags}. In particular, both the critical lines hosting unnecessary criticality and multiversality can be reached by tuning a single parameter with no additional fine-tuning.  
	
	\medskip 
	
	\noindent\emph{Discussion:} We conclude by sketching conditions to generate  models with multiversality and unnecessary criticality. It is illuminating to anchor the discussion to the region on the critical surface where the universality class is about to change or the surface is about terminate. Broadly, we need two ingredients (i) a single parameter $\delta$ that couples to all relevant operators that lead to the gapped phase(s) and (ii) a marginal operator $\mathcal{O}_M$ whose energy can be minimized simultaneously with that of the operators coupled to $\delta$. The change along the critical surface occurs when $\mathcal{O}_M$ changes from marginally irrelevant to marginally relevant. For the examples in this work, $\delta$ couples to $\mathcal{U}_{1,2}$ and $\mathcal{O}_M\propto  \mathcal{V}_\pm$ in the bosonized language. These conditions are neither necessary nor sufficient, but are useful guides. Two additional ingredients also serve to simplify our analysis. The first is the existence of an exactly marginal operator $\partial_x \phi_1\partial_x \phi_2$ that can tune the scaling dimension of $\mathcal{O}_M$ along the critical surface. The second is the  ``failed SPT" premise~\cite{AnfusoRosch_PhysRevB.75.144420,BiSenthil_PhysRevX.9.021034,JianXu_PhysRevB.101.035118,ThorngrenVishwanathVerresen_IntrinsicallyGapless_PhysRevB.104.075132,Pollmann_Quotient} which provides a template to construct phase diagrams using  results from the classification of SPT phases~\cite{ChenGuWen_1DSPT_PhysRevB.84.235128,supp}. Using similar ingredients, it is likely that one can engineer examples of both phenomena in higher dimensions. 
	
	Finally, we flag some  possible extensions of this work. First, note that our two-leg models can be straightforwardly generalized to $2N$ legs where the possible critical phenomena are richer~\cite{supp}. Second, note that we restricted our focus to multiversality on the critical surface of a phase transition separating a trivial from a nontrivial SPT phase, which lies outside the Landau paradigm of symmetry-breaking orders. It would be equally interesting to find examples where the transition is {\it not} Landau forbidden, but one of the multiversality classes is~\cite{LandauBeyondLandau_PhysRevResearch.2.023031,supp}. 
	Third, we conjectured that unnecessary criticality, boundary criticality, and stable boundary modes are intimately connected. It would be useful to make this more concrete, e.g. via a field-theoretic formulation, particularly in higher dimensions. A fourth open question is whether  phase diagrams analogous to those studied in this Letter can be obtained in models with quenched randomness. Finally, it would be particularly exciting to find experimental examples of either of these phenomena. Given the simplicity of the models presented here, we are optimistic that this is a question that can be answered positively in the not-too-distant future.

	\medskip 
	
	\noindent\emph{Note added:} In the final stages of the preparation of this manuscript, we became aware of an upcoming independent work~\cite{toappear} which also studies unnecessary criticality in spin chains. We thank the authors for alerting us about their results.
	
	\medskip 
	\noindent\emph{Acknowledgments}: We thank Sounak Biswas, Yizhi You, Fabian Essler, Nick Bultinck, Paul Fendley, Mike Blake, Senthil Todadri, Apoorv Tiwari, Dan Arovas, Soonwon Choi, and Sam Garratt for helpful discussions and correspondence. The numerical analysis in this work was performed using the ITensor library~\cite{itensor}. We acknowledge support from the  European Research Council under the European Union Horizon 2020 Research and Innovation Programme, Grant Agreement No. 804213-TMCS.

	
	
	\bibliography{references}{}
	
	\ifarXiv
	

\section*{Supplementary material}

# Supplementary material for
# "Multiversality and unnecessary criticality in one dimension"

Abhishodh Prakash,[1, *] Michele Fava,[1, †] and S.A. Parameswaran[1, ‡]

[1]*Rudolf Peierls Centre for Theoretical Physics, University of Oxford, Oxford OX1 3PU, United Kingdom*

## CONTENTS

## I. CONFIRMING THE PHASE DIAGRAM OF $H^M$

We now provide additional details to confirm the phase diagrams shown in the main text for the two-leg spin ladder models.

---

* abhishodh.prakash@physics.ox.ac.uk (he/him/his)
† michele.fava@physics.ox.ac.uk
‡ sid.parameswaran@physics.ox.ac.uk

## A. Absence of additional critical lines in the phase diagram of $H^M$

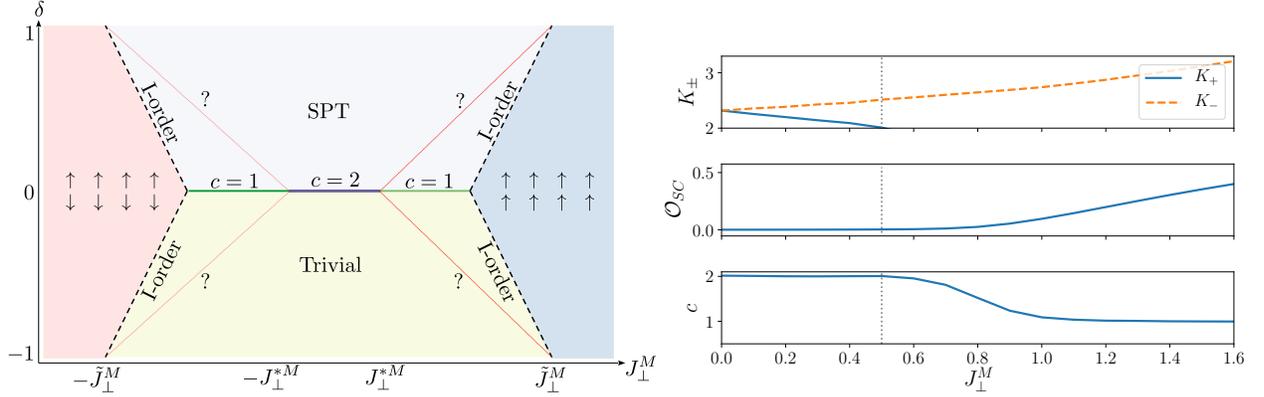

FIG. S1. **Left:** Schematic phase diagram of $H^M$. The lines in marked with a question mark would eliminate multiversality and turn the phase diagram into a conventional one if present as critical lines. **Right:** The Luttinger parameters $K_\pm$, string order parameter $\mathcal{O}_{SC}$ defined in the main text and central charge for various $J_\perp^M > 0$ along the $\delta = 0$ line also shown in the main text. The vertical line marks $J_\perp^{*M} \approx 0.6$ where the central charge changes from $c = 2$ to $c = 1$.

Let us begin with $H^M$ which is defined as

$$H^M = \sum_{j=1}^L \sum_{\alpha=1}^2 \left(1 + (-1)^j \delta\right)\left[S_{\alpha j}^x S_{\alpha j+1}^x + S_{\alpha j}^y S_{\alpha j+1}^y + \Delta S_{\alpha j}^z S_{\alpha j+1}^z\right] + J_\perp^M \sum_j S_{1j}^z S_{2j}^z, \tag{1}$$

with fixed $\Delta \in \left(-\frac{1}{\sqrt{2}}, 0\right)$ (we take $\Delta = -0.25$ in all our numerical studies). The phase diagram for $H^M$, shown in the main text, is reproduced in fig. S1 (left) for convenience. First, we consider the possibility that we are missing additional critical lines in the phase diagram (marked by dotted red lines labeled with a question mark fig. S1) which, if present, would eliminate multiversality leaving a more conventional phase diagram. This is not the case, as can be determined in two ways. First, the value of the Luttinger parameters extracted numerically (fig. S1 (right)) shows that there are no relevant operators allowed by symmetry apart from those considered in the main text i.e. $\mathcal{U}_1 + \mathcal{U}_2 \equiv \cos\phi_1 + \cos\phi_2$ and $\mathcal{V}_\pm \equiv \cos(\phi_1 \pm \phi_2)$. For example, the operator with the next smallest scaling dimensions,

$$[\cos(2\phi_1) + \cos(2\phi_2)] = \begin{cases} K_+ & \text{for } J_\perp^M < -J_\perp^{*M} \\ K_+ + K_- & \text{for } |J_\perp^M| < |J_\perp^{*M}| \\ K_- & \text{for } J_\perp^M > J_\perp^{*M} \end{cases} \tag{2}$$

is always irrelevant (recall that $J_\perp^{*M}$ labels the value of $J_\perp^M$ where the central charge changes from $c = 2$ to $c = 1$ on the $\delta = 0$ line). Therefore, we do not expect any additional phases or transitions. In the rest of this subsection, we will also also verify this numerically.

Before doing so, first we argue that the only shape the putative additional critical lines can take has the topology shown in fig. S1 (left). First, observe that $\delta \mapsto -\delta$ and $J_\perp^M \mapsto -J_\perp^M$ are related by unitary transformations and therefore, the phase diagram is expected to be symmetric about the $\delta$ and $J_\perp^M$ axes. Next, for the limit of $\delta = \pm 1$, eq. (1) is exactly solvable as it splits up into local terms with support on four spins (let us consider periodic boundary conditions to keep things simple). The local four-spin Hamiltonian is

$$h(\Delta, J_\perp^M) = 2 \sum_{\alpha=1}^2 \left(S_{\alpha j}^x S_{\alpha j+1}^x + S_{\alpha j}^y S_{\alpha j+1}^y + \Delta S_{\alpha j}^z S_{\alpha j+1}^z\right) + J_\perp^M \left(S_{1j}^z S_{2j}^z + S_{1j+1}^z S_{2j+1}^z\right). \tag{3}$$

The phase diagram in this limit can be determined by diagonalizing eq. (3) where there exists a *single* level-crossing between a unique ground state (representing the trivial and SPT phases in the phase diagram of fig. S1) to a degenerate ground state (representing the magnetic phases in the phase diagram of fig. S1) at $\pm \tilde{J}_\perp^M = \pm \frac{(1-\Delta^2)}{|\Delta|}$. This is interpolated to the full phase diagram shown in the main text as a single first-order line separating the magnetic phases from the trivial and SPT phases. Thus, the concrete alternative we wish to rule out is that the points

$\left(\delta = \pm 1, \pm \tilde{J}_\perp\right)$ could extend into the phase diagram through the multiple critical lines as shown in the left panel of fig. S1 (marked with a question mark).

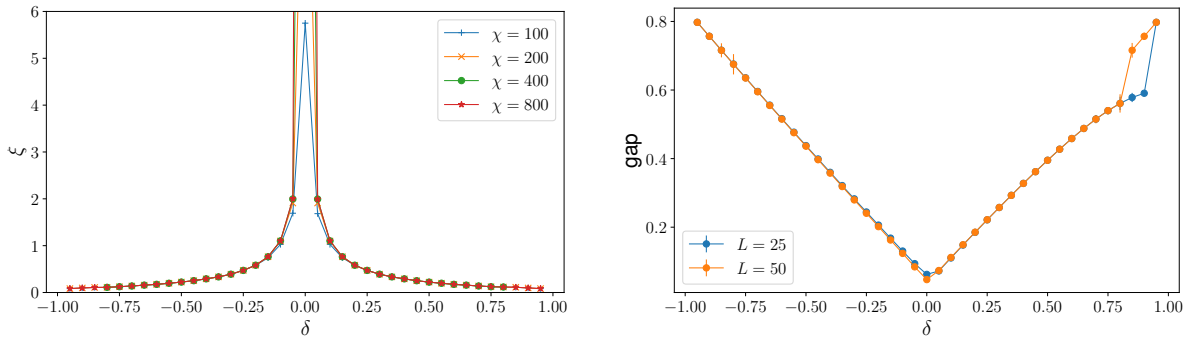

FIG. S2. **Left:** correlation length $\xi$ of $H^M$ for various $\delta$ across a cut at $J_\perp^M = 1.2 > J_\perp^{*M}$ which corresponds to a $c = 1$ region for $\delta = 0$. **Right:** energy gap between the ground state, lying in the $(\sum_j S_{1j}^z, \sum_j S_{2j}^z) = (0,0)$ sector, and the lowest energy state in the $(\sum_j S_{1j}^z, \sum_j S_{2j}^z) = (1,0)$ sector. $L$ denotes the number of unit cells in the system. The jump of the estimated gap for $\delta \gtrsim 0.75$ and $L = 50$ is due to a metastability of the first excited state in the DMRG algorithm. The numerical plots are performed for $H^M$ in eq. (1) with fixed $\Delta = -0.25$. These plots tell us that there are no additional critical points apart from $\delta = 0$ and therefore the lines marked with a question mark in fig. S1 do not exist.

To rule out this possibility numerically, we compute the correlation length $\xi$ for fixed $J_\perp^M = 1.2 > J_\perp^{*M}$ that cuts through the $c = 1$ region on the $\delta = 0$ line. $\xi(\chi)$ can be directly computed as the second-largest eigenvalue of the transfer matrix of the matrix-produce-state (MPS) [1]. As shown in fig. S2 (left), within the accuracy of our numerical analysis, we can rule out the possibility of additional second-order phase transitions for $\delta \neq 0$ since the correlation length $\xi$ remains finite throughout this line except at $\delta = 0$. Furthermore, we rule out the possibility of a first-order phase transition by studying the energy gap between the zero-magnetization $(\sum_j S_{1j}^z, \sum_j S_{2j}^z) = (0,0)$ sector, where the expected ground state lies, and the $(\sum_j S_{1j}^z, \sum_j S_{2j}^z) = (1,0)$ sector. We show in fig. S1 (right) that this gap always remains finite away from $\delta = 0$.

### B. Distinguishing the trivial and non-trivial SPT phases using string order parameters

We now provide details of how the trivial and non-trivial SPT phases are distinguished using a string order parameter [2]. Since the non-trivial SPT phase in eq. (1) descends from two copies of the dimerized XXZ chain, we can use the same string order parameters that can distinguish between the SPT phases in a single chain. As usual, we can

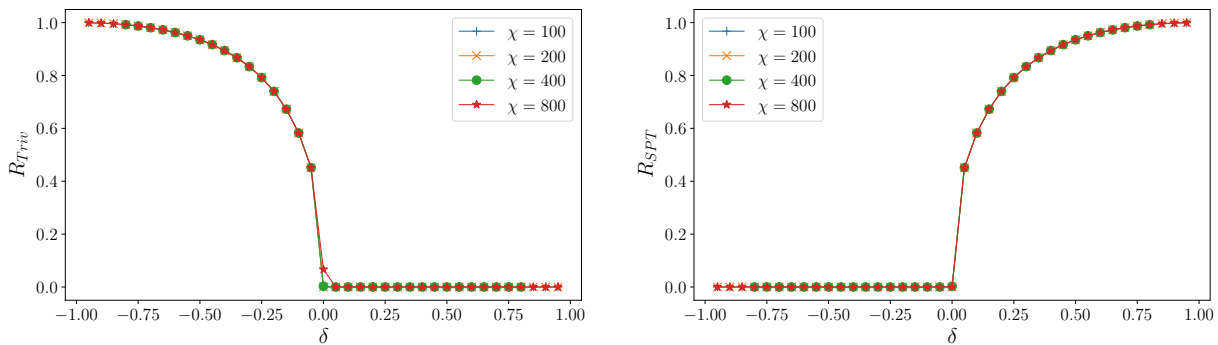

FIG. S3. String order parameters $R_{Triv}^1 = R_{Triv}^2 = R_{Triv}$ and $R_{SPT}^1 = R_{SPT}^2 = R_{SPT}$, defined in eqs. (4) and (5) as a function of $\delta$ for $H^M$ at $J_\perp^M = 1.2$ and $\Delta = -0.25$. Curves at different bond dimensions are meant to show that the numerical evaluation is converged in bond dimension.

consider the following two string order parameters:

$$R_{Triv}^{\alpha} = \lim_{r\to\infty} \left\langle \sigma^z_{\alpha,2j+1}\sigma^z_{\alpha,2j+2} \left(\prod_{j'=j}^{j+r} \sigma^z_{\alpha,2j'+1}\sigma^z_{\alpha,2j'+2}\right) \sigma^z_{\alpha,2j+2r+1}\sigma^z_{\alpha,2j+2r+2} \right\rangle \quad (4)$$

$$R_{SPT}^{\alpha} = \lim_{r\to\infty} \left\langle \sigma^z_{\alpha,2j+2} \left(\prod_{j'=j}^{j+r} \sigma^z_{\alpha,2j'+1}\sigma^z_{\alpha,2j'+2}\right) \sigma^z_{\alpha,2j+2r+1} \right\rangle. \quad (5)$$

where $\vec{\sigma}_{\alpha j}$ are Pauli matrices $\vec{\sigma}_{\alpha j} = 2\vec{S}_{\alpha j}$. $R_{Triv}^{\alpha}$ and $R_{SPT}^{\alpha}$ pick up non-zero expectation values for the trivial and non-trivial SPT phases respectively. $\alpha$ labels the two legs of the spin ladders and we can pick any one for our purposes. Figure S3 shows the string order parameters eqs. (4) and (5) evaluated on the ground state for $J_\perp^M = 1.2$ where we see that they can sharply distinguish between the trivial and non-trivial SPT phases.

## II. CONFIRMING THE PHASE DIAGRAM OF $H^U$

In this section we provide additional numerical data confirming the phase diagram of $H^U$ shown in the main text and also reproduced in the left panel of fig. S4 for convenience.

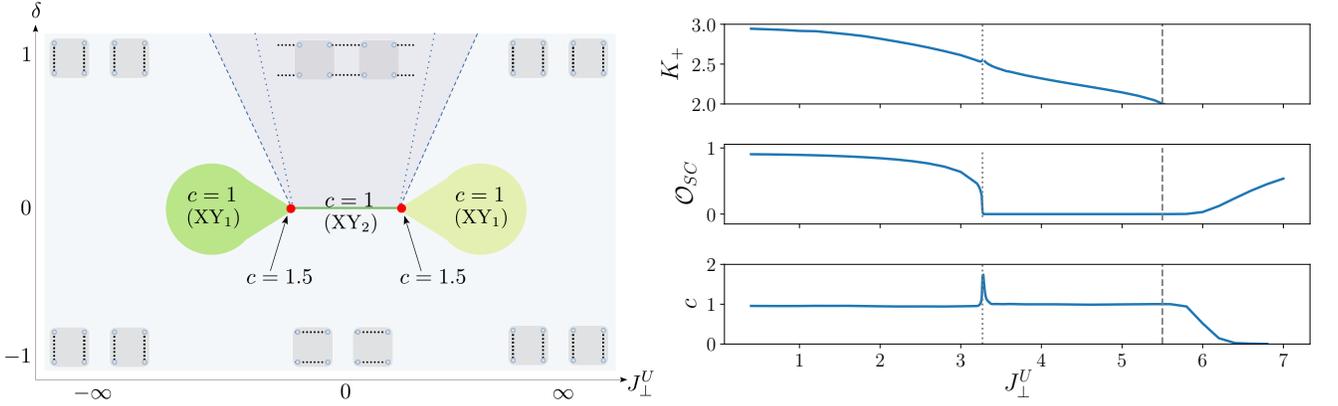

FIG. S4. **Left:** Schematic phase diagram of $H^U$. The shaded region above the XY$_2$ line denotes the presence of stable two-fold degenerate boundary modes. The dashed line surrounding the shaded region indicates the boundary transition where the boundary modes eliminated. The dotted line represents an entanglement transition where the dominant value of the entanglement spectrum changes from two-fold degenerate to unique. **Right:** (from top to bottom) Luttinger parameter $K_+$, string order parameter $\mathcal{O}_{SC}$ and central charge $c$ along the $\delta = 0$ line of $H^U$ for various $J_\perp^U > 0$ with fixed $\Delta = -0.05$ and $J_\perp^M = -5.2$. The dotted line ($J_\perp^U \approx 3.27$) denotes the $c = \frac{3}{2}$ point when the XY$_2$ critical line transitions to XY$_1$ lobes. The dashed line ($J_\perp^U \approx 5.5$) denotes the point where $K_+ = 2$ where the system transitions to a trivial gapped phase.

### A. Effect of $\delta \neq 0$

Let us begin with $H^U$ which is defined as

$$H_\perp^U = \sum_{j=1}^L \sum_{\alpha=1}^2 \left(1+(-1)^j\delta\right)\left[S^x_{\alpha j}S^x_{\alpha j+1} + S^y_{\alpha j}S^y_{\alpha j+1} + \Delta S^z_{\alpha j}S^z_{\alpha j+1}\right] + J_\perp^M \sum_j S^z_{1j}S^z_{2j} + J_\perp^U \sum_j \left(S^x_{1j}S^x_{2j} + S^y_{1j}S^y_{2j}\right), \quad (6)$$

with $\Delta$ and $J_\perp^M$ fixed to some value such that, when $\delta = J_U = 0$ the Hamiltonian correspond to a $c = 1$ point with pinned $\phi_-$ in the multiversality phase diagram (we fix $\Delta = -0.05$ and $J_\perp^M = -5.2$ for all our numerical analysis).

In the main text, we argued that for $J_U \lesssim 2.27$ and $\delta = 0$, the system is described by a XY$_2$ Luttinger liquid where $\phi_1 - \phi_2$ is pinned, while $\phi_1 + \phi_2$ is not subject to any relevant perturbations, thus giving rise to a $c = 1$ CFT. In this case, $\delta$ constitutes a relevant perturbation, coupling to an operator with scaling dimension $K_+/4$. Therefore, the system becomes immediately gapped for any $\delta \neq 0$. We confirm this numerically by studying the correlation length

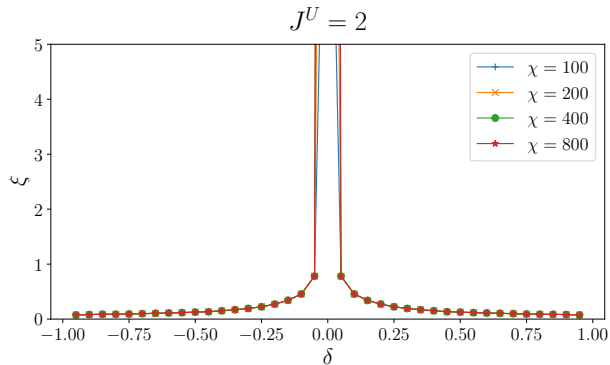

FIG. S5. **Left:** Correlation length $\xi$ as a function of $\delta$ for $J_\perp^U = 2$. $\xi$ appears to be finite for any $\delta > 0$, in agreement with the phase diagram S4 for the parameter region where we have an XY$_2$ unnecessary critical line.

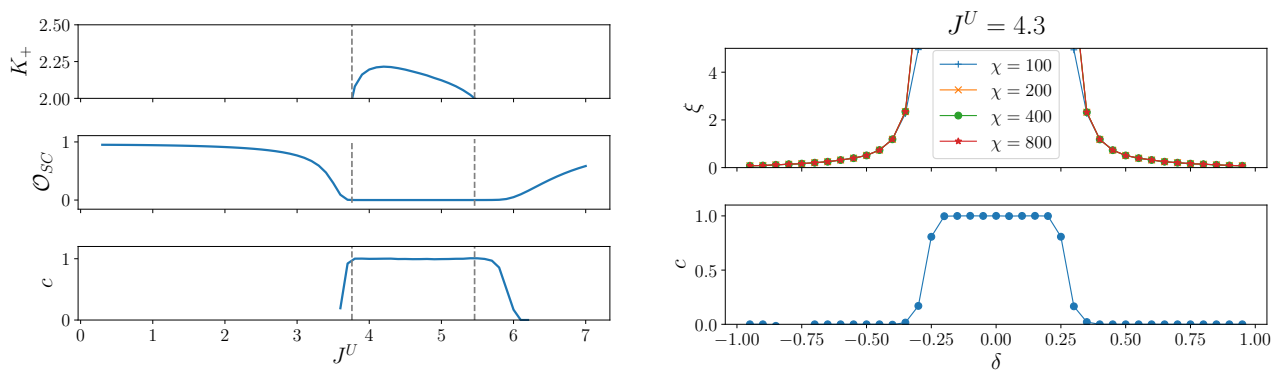

FIG. S6. **Left:** (from top to bottom) Luttinger parameter $K_+$, string order parameter $\mathcal{O}_{\mathrm{SC}}$ and central charge $c$ along the $\delta = 0.1$ line of $H^U$ for various $J_\perp^U > 0$. The dashed lines ($J_\perp^U \approx 3.76$ and $J_\perp^U \approx 5.46$) denote the transition points when the XY$_2$ critical region gaps out, in correspondence of the points where $K_+ = 2$. **Right:** Correlation length and central charge as a function of $\delta$ for $J_\perp^U = 4.3$. The correlation length is finite when $|\delta| \gtrsim 0.25$, whereas, for $|\delta| \lesssim 0.25$ finite-entanglement scaling is compatible with $c = 1$. Both plots are for fixed $\Delta = -0.05$ and $J_\perp^M = -5.2$. The two figures together are consistent with the phase diagram shown in the left panel of fig. S4.

as a function of $\delta$ in fig. S5 for a fixed $J_\perp^U = 2$. Indeed, we see that the correlation length attains a finite value already for $|\delta| = 0.05$, giving strong evidence that the gapless XY$_2$ region is a critical line.

Conversely, for $2.27 \lesssim J_U \lesssim 5.5$ and $\delta = 0$ we argued in the main text that the system is described by a XY$_1$ Luttinger liquid phase where $\theta_1 - \theta_2$ is pinned, while the field $\phi_1 + \phi_2$ is again not subject to any relevant perturbations and gives rise to a $c = 1$ CFT. In this case $\mathcal{U}_{1,2}$ are not a scaling operators: their two-point functions decay exponentially over a finite correlation length $\xi_-$. We then expect $\delta(\mathcal{U}_1 + \mathcal{U}_2)$ to be an irrelevant perturbation and the system to remain gapless up to finite $\delta$, thus producing the gapless area in the phase diagram of fig. S4. We confirm this by fixing $J_\perp^U = 4.3$ and studying the correlation length $\xi$ as a function of $\delta$ in the right panel of fig. S6. In agreement with our expectations, the system appears to be described by a $c = 1$ theory for $-0.25 \lesssim \delta \lesssim 0.25$, whereas the correlation length $\xi$ is finite outside this range. As a further confirmation of the phase diagram at finite $\delta$, we can similarly study a cut through the phase diagram along the line with $\delta = 0.1$. As the left panel of fig. S6 shows, $H_U$ is gapped for small values of $J_\perp^U$, and becomes gapless for $3.76 \lesssim J_\perp^U \lesssim 5.46$. We show that this region —whose edges are marked by dashed lines in Fig. S6— is compatible with $c = 1$ and can be further identified by the condition $K_+ > 2$ or, alternatively, $\mathcal{O}_{SC} = 0$, giving strong evidence in favour of the phase diagram shown in fig. S4.

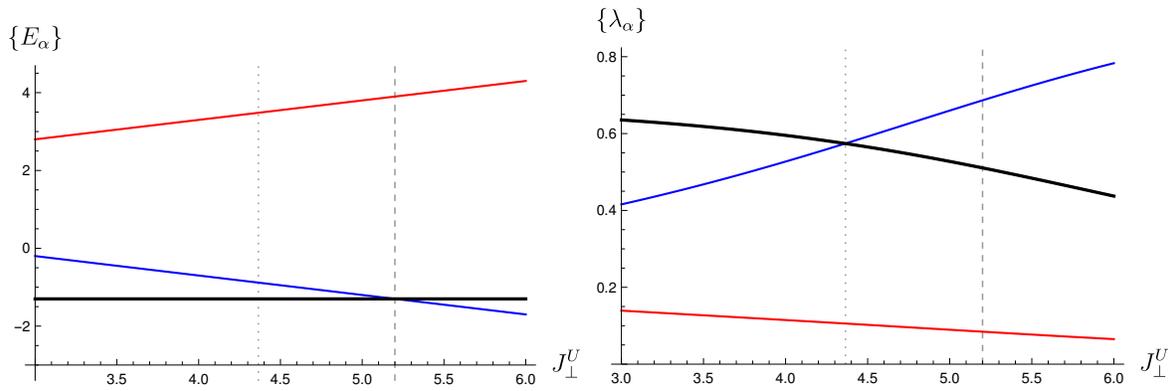

FIG. S7. Left: Spectrum of the effective boundary Hamiltonian of $H^U$ for $\delta = 1$ shown in eq. (7) where we see a transition at $(|J_\perp^U| = |J_\perp^M|)$. Right: Schmidt values for the bipartite Schmidt decomposition of the ground state of the effective bond Hamiltonian shown in eq. (8). We used the same parameters used for numerical analysis– $J_\perp^M = -5.2$ and $\Delta = -0.05$. The solid black curve for both plots represents a level with two-fold degeneracy. Vertical lines represent boundary (dashed) and entanglement (dotted) transitions

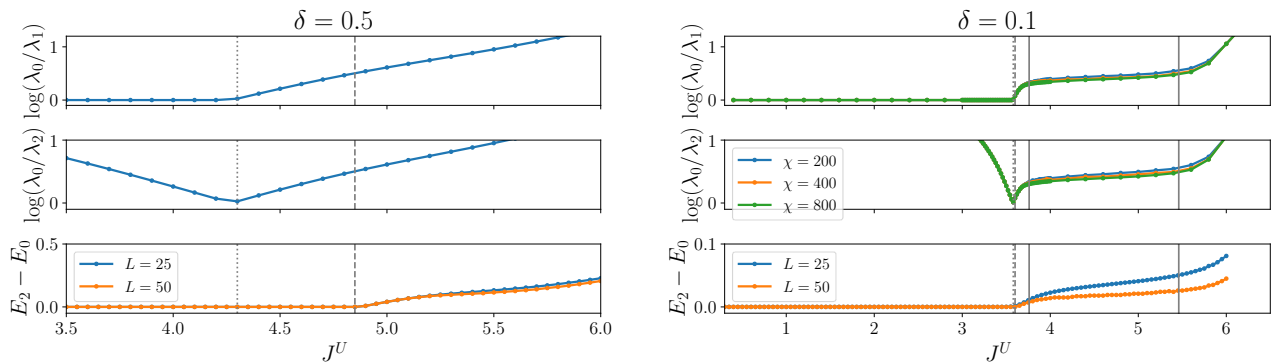

FIG. S8. From top to bottom: $\log(\lambda_0/\lambda_1)$, $\log(\lambda_0/\lambda_2)$ denoting the gap in the entanglement spectrum, and the energy gap between the $S^z = 0$ and $S^z = 2$ sector for $\delta = 0.5$ (left) and $\delta = 0.1$ (right). Here $\lambda_0$, $\lambda_1$ and $\lambda_2$ are the three highest Schmidt values for a cut between two unit cells and $E_2$ and $E_0$ are the ground state energies in the $S^z = 0$ and $S^z = 2$ symmetry sectors. In the left panel, the dashed and dotted lines represent boundary and entanglement transitions at $J_\perp^U \approx 4.85$ and $J_\perp^U \approx 4.3$ respectively. In the right panel, the boundary transition (dashed line) occurs at $J_\perp^U \approx 3.6$ and entanglement transition (dotted line) occurs at $J_\perp^U \approx 3.58$. The two additional solid lines denote the boundary of the $XY_1$ region, as determined in Fig. S6(left).

### B. Boundary and entanglement transition for $\delta > 0$

Finally, we turn our attention to the region of the phase diagram of $H^U$ with boundary modes. As discussed in the main text, this can be analyzed analytically in the limiting case of $\delta = 1$ where the effective boundary Hamiltonian on each end can be described as

$$H_\partial = J_\perp^M S_1^z S_2^z + J_\perp^U \left(S_1^x S_2^x + S_1^y S_2^y\right). \tag{7}$$

By diagonalizing eq. (7), as shown in the left panel of fig. S7, we can see that the region with two-fold degenerate boundary modes ($|J_\perp^U| > |J_\perp^M|$) terminates at $(|J_\perp^U| = |J_\perp^M|)$ when the system undergoes a boundary transition where boundary modes become three-fold degenerate. Here we provide numerical evidence that the region with boundary modes persists for other values of $\delta$ and the boundary transition merges with the bulk transition at the $c = 3/2$ point, as highlighted in fig. S4 in the form of dotted lines on the boundary of the shaded region.

The boundary modes can be detected numerically by computing the lowest energy states in distinct symmetry sectors where the boundary modes live– $S^z = 0, \pm 2$ as determined from diagonalizing eq. (7) where $S^z \equiv \sum_j S_{1j}^z + \sum_j S_{2j}^z$. So long as the boundary modes exist, the ground state in these sectors are degenerate. When the boundary modes are eliminated, only the $S^z = 0$ sector contains the unique ground state of the system. The bottom panels of fig. S8

shows the difference between the ground state energies of the $S^z = 0$ and $S^z = +2$ sectors away from the exactly solvable limit, for $\delta = 0.5$ (left) and $\delta = 0.1$ (right). We indeed see that the boundary modes persist and approach the $c = \frac{3}{2}$ point as $\delta$ is reduced.

A useful bulk probe that serves as a proxy for boundary modes is the entanglement spectrum [3–6] corresponding to the Schmidt values $\{\lambda_\alpha\}$ for a Schmidt decomposition across a cut between two unit cells. Note that we have implicitly defined a unit cell to contain four qubits enclosing lattice points labelled $\{(1j), (1, j+1), (2j), (2j+1)\}$ where $j$ is odd. The presence of boundary modes is signalled by the degeneracy of the largest Schmidt value. In the limiting case of $\delta = 1$, this can be obtained from the following effective 4-qubit Hamiltonian across a bond between two unit cells

$$H_{bond} = J_\perp^M \left(S_{1j}^z S_{2j}^z + S_{1j+1}^z S_{2j+1}^z\right) + J_\perp^U \left(S_{1j}^x S_{2j}^x + S_{1j}^y S_{2j}^y + S_{1j+1}^x S_{2j+1}^x + S_{1j+1}^y S_{2j+1}^y\right)$$
$$+ 2 \sum_{\alpha=1}^{2} \left(S_{\alpha j}^x S_{\alpha j+1}^x + S_{\alpha j}^y S_{\alpha j+1}^y + \Delta S_{\alpha j}^z S_{\alpha j+1}^z\right). \quad (8)$$

As shown in fig. S7, by tracking $\{\lambda_\alpha\}$ for a Schmidt decomposition between two unit cells, in the regime of interest studied in the main text, we see that for small $|J_\perp^U|$ the largest Schmidt value is doubly-degenerate and for large $|J_\perp^U|$, it is unique. These two regimes are separated by an entanglement transition where the leading Schmidt value is triply degenerate. This is consistent with the expectation [3] that the entanglement spectrum reflects the nature of the low-lying boundary spectrum. Curiously however, the entanglement transition occurs at a value $|J_\perp^U| < |J_\perp^M|$ which is different from (and smaller than) the boundary transition, similar to the phase diagram of [7].

For $\delta < 1$ i.e. away from the exactly solvable limit, by tracking the three leading Schmidt values $(\lambda_0, \lambda_1, \lambda_2)$ we can verify numerically that the same picture qualitatively holds. In the top two panels of both plots in fig. S8, we report $\log(\lambda_0/\lambda_1)$ and $\log(\lambda_0/\lambda_2)$ as a function of $J_\perp^U$ for $\delta = 0.5$ (left) and $\delta = 0.1$ (right). For $\delta = 0.5$, we see qualitatively the same picture as for $\delta = 1$ although the separation between the boundary and entanglement transition has reduced. For $\delta = 0.1$, we can distinguish four regions from the view of the entanglement spectrum. Starting from $J_\perp^U = 0$ and increasing its value, we first encounter a gapped region where $\lambda_1 = \lambda_0$, signalling the presence of edge modes. At $J_\perp^U \approx 3.58$, we have $\lambda_2 = \lambda_1 = \lambda_0$ denoting the entanglement transition. This is very close to the boundary transition ($J_\perp^U \approx 3.6$) At larger values of $J_\perp^U$ ($3.76 \lesssim J_\perp^U \lesssim 5.46$), the system enters the $XY_1$ region. Here the gap of the entanglement spectrum is finite for any finite bond dimension $\chi$, but tends to zero in the $\chi \to \infty$ limit, as expected for critical theories. Finally, for $J_\perp^U \gtrsim 5.46$, the system enters a gapped phase without edge modes. Both the boundary and entanglement transitions are expected to get progressively closer to the bulk transition to the $XY_1$ phase with decreasing values of $|\delta|$ and meet when $\delta = 0$ at the $c = 3/2$ point (at $J_\perp^U \approx 3.27$) separating the $XY_1$ and $XY_2$ regions. Note that all the observations above are consistent with the proposed phase diagram shown in the left panel of fig. S4.

### III. FURTHER NUMERICAL DATA

#### A. Extraction of the Luttinger parameter from iDMRG

In this section we provide additional details on the estimation of the Luttinger parameters $K_\pm$ reported in Fig. 3 and 4 of the main text. We do this by computing the two-point correlation functions of suitable lattice operators whose scaling form can be obtained from bosonization. The lattice operators we use and their bosonized forms are as follows

$$S_1^+ S_2^+ \sim e^{i(\theta_1 + \theta_2)}, \ S_1^+ S_2^- \sim e^{i(\theta_1 - \theta_2)}. \quad (9)$$

The form of their two-point correlation functions can be written as

$$C_+ = \langle \left(S_{1j}^+ S_{2j}^+\right)\left(S_{1j+r}^- S_{2j+r}^-\right)\rangle \sim r^{-2/K_+}, \quad (10)$$
$$C_- = \langle \left(S_{1j}^+ S_{2j}^-\right)\left(S_{1j+r}^- S_{2j+r}^+\right)\rangle \sim r^{-2/K_-}. \quad (11)$$

By numerically computing $C_\pm$ using iDMRG, we can extract $K_\pm$. The behaviour of $C_\pm$ for some representative values of $J_\perp^M$ of the Hamiltonian $H^M$ (eq. (1)) is reported in fig. S9 (top row). When the $\pm$ sectors are gapless the correlators $C_\pm$ are compatible with power-law scaling when $r$ is large enough that the asymptotic form in the equation above holds, while remaining smaller than the correlation length $\xi(\chi)$ introduced by the finite bond dimension employed in the iDMRG simulations. In practice we find that for most of the bond dimensions we used, there is a good agreement

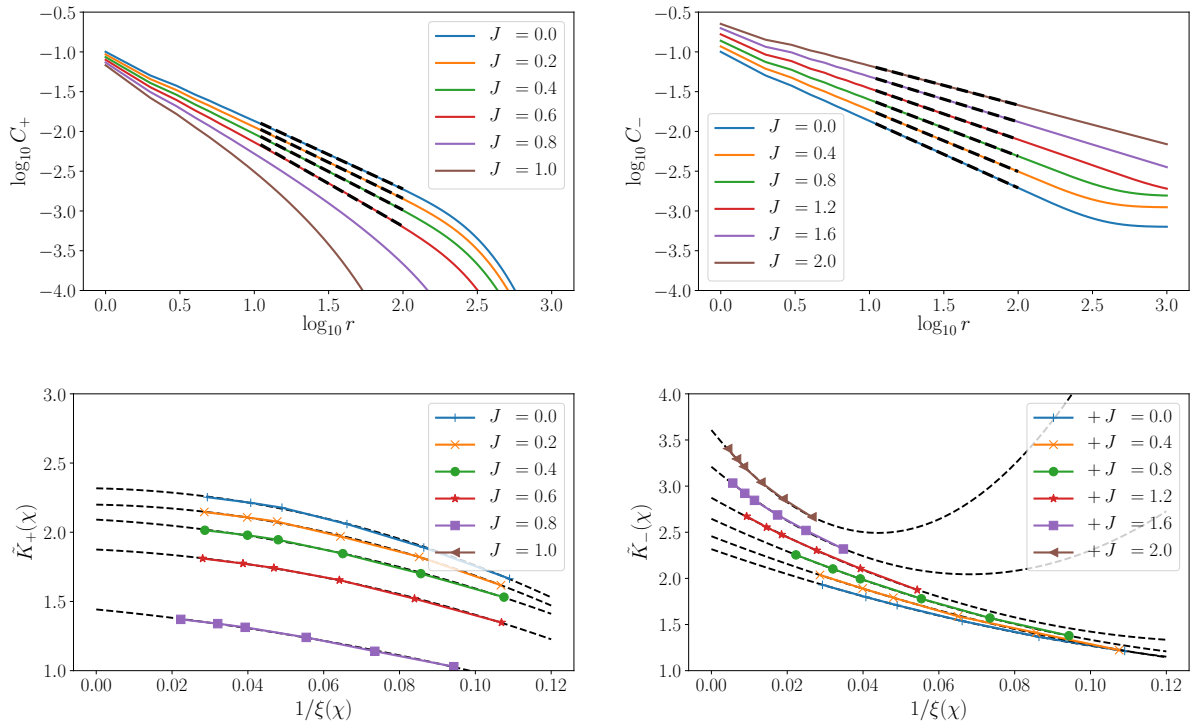

FIG. S9. **Top:** Examples of the behaviour of $C_\pm$ shown in eqs. (10) and (11) for some representative values of $J_\perp^M$ of the Hamiltonian $H^M$ (eq. (1)). Coloured continuous lines denote the iDMRG data for bond dimension $\chi = 1500$. Black dashed lines in the interval $r \in [5, 50]$ are used for the fit to a power law. Note that for $J_\perp^M \gtrsim 0.5$ the + sector is gapped and therefore we expect $C_+$ to ultimately cross over to an exponential decay at large distances. Nonetheless, for $J_\perp^M \lesssim 0.7$ we can still define an approximate $K_+$ by fitting $C_+$ to a power law in the crossover region. **Bottom:** Examples of the extrapolations of $\tilde{K}_\pm(\chi)$ (eq. (12)) to the $\chi \to \infty$ limit, which coincides with the $\xi(\chi) \to \infty$ limit. In the figure different points denote different bond dimensions $\chi \in \{300, 400, 550, 800, 1000, 1500\}$.

with a power-law scaling for $5 \leq r \leq 50$ (see fig. S9, top row). Therefore by fitting $C_\pm$ in this interval to the power-law form shown in eqs. (10) and (11) we obtain $\tilde{K}_\pm(\chi)$ as an estimate of the true $K_\pm$ for bond dimension $\chi$.

Note that, even if the correlation length for a given bond dimension $\chi$ is much larger than the length of the interval $r$ used to fit the power-law scaling, the estimates $\tilde{K}_\pm(\chi)$ still has a spurious dependence of $\chi$, and the exact values $K_\pm$ are recovered only in the $\chi \to \infty$ limit. Since at the maximum bond dimensions we can access the estimate of $\tilde{K}_\pm(\chi)$ are not converged, we can extrapolate our results to the $\chi \to \infty$, by fitting the $\chi$ dependence as

$$\tilde{K}_\pm(\chi) = K_\pm + \frac{A_\pm}{\xi(\chi)} + \frac{B_\pm}{\xi^2(\chi)}. \quad (12)$$

Here $\xi(\chi)$ can be directly computed as the second-largest eigenvalue of the transfer matrix of the MPS. We report the data used for the extrapolation in fig. S9 (bottom row) for some representative values in the multiversality case.

The same fitting and extrapolation procedures has been used to determine $K_+$ in $H^U$ (eq. (6)). Sample plots for representative $J_\perp^U$ are shown in fig. S10.

### B. Convergence of the central charge

In this subsection we present further data related to the convergence in bond dimension of the central charge estimates for $H_M$ and $H_U$ reported in Fig. S1 and Fig. S4 respectively. We ran numerical simulations for $H^M$ for the following bond dimensions $\chi \in \{150, 200, 300, 400, 550, 800, 1000, 1500\}$. Similarly, we ran simulations for $H^U$ with $\chi \in \{150, 200, 300, 400, 550, 800\}$. For each bond dimension we estimated the correlation length of the MPS $\xi(\chi)$ and its (Von Neumann) entanglement entropy $S(\chi)$. For a Hamiltonian whose low-energy properties are decribed by a conformal field theory, both $\xi(\chi)$ and $S(\chi)$ are expected to grow as a function of $\chi$. Furthermore, they are related

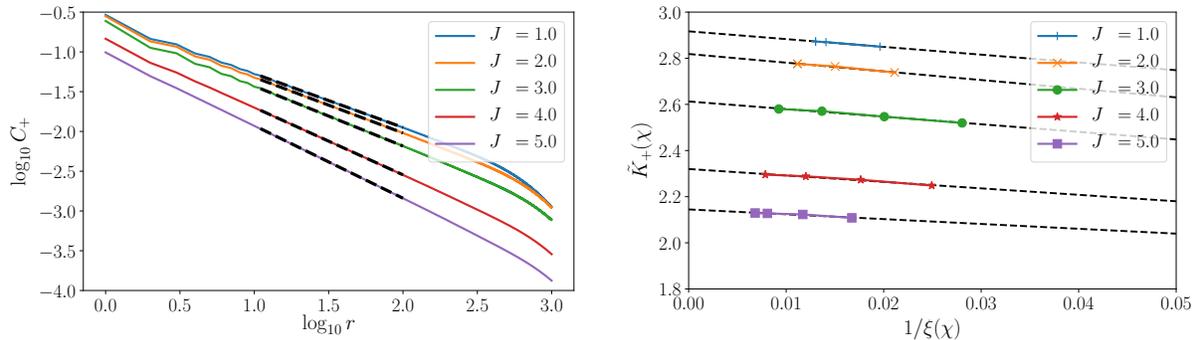

FIG. S10. (left panel) Examples of the behaviour of $C_\pm$ for some representative values of $J_\perp^U$ in the unnecessary criticality case. Colored continuous lines denote the iDMRG data for bond dimension $\chi = 800$. Black dashed lines show the fit to a power law with $r$ in the interval $r \in [5, 50]$. (right panel) Examples of the extrapolations of $\tilde{K}_+(\chi)$ to the $\chi \to \infty$ limit, which coincides with the $\xi(\chi) \to \infty$ limit. In the figure different points denote different bond dimensions $\chi \in \{300, 400, 550, 800\}$.

by [8–10]

$$S(\chi) \sim \frac{c}{6} \log \xi(\chi). \tag{13}$$

The central charge $c$ can then be estimated through a linear fit. Fig. S1 and Fig. S4 have been obtained by using data from all bond dimensions above. To test the convergence of $c$ we performed the finite-entanglement-scaling fit using only bond dimensions within a variable range $[\chi_{\min}, \chi_{\max}]$, as reported in Fig. S11 (middle and right panel).

For concreteness we begin by discussing the estimate of $c$ for $H^M$. Here we see that the central charge estimate is well converged in bond dimension away from the transition point at $J_\perp^M = J_\perp^{*M} \approx 0.6$. Instead, near the transition point, $c$ is supposed to have a discontinuous jump from $c = 2$ to $c = 1$ as $J_\perp^M$ is increased. This discontinuous jump is smoothed out in numerical simulation as an effect of the finite bond dimension employed and is supposed to be recovered only in the limit of $\chi \to \infty$. The expected behaviour is sketched in the leftmost panel of Fig. S11: as the bond dimension $\chi$ is increased the jump becomes sharper and drifts towards the actual transition point. The underlying intuition is that at the right of the transition point there is a crossover lengthscale $\xi_{\mathrm{cr}}$, such that running the RG up to lengthscale $\xi_{\mathrm{cr}}$ the relevant coupling driving the transition has not grown to be $O(1)$ yet. According to BKT scaling this length scales as [11]

$$\xi_{\mathrm{cr}} \sim \exp\left(\frac{\mathrm{const.}}{\sqrt{J_\perp^M - J_\perp^{*M}}}\right) \tag{14}$$

Therefore the state is effectively described by a CFT with $c = 2$ up to lengthscales $\xi_{\mathrm{cr}}$. For the entanglement scaling this means that

$$S(\chi) \sim \begin{cases} \frac{1}{3} \log \xi(\chi), & \xi(\chi) \lesssim \xi_{\mathrm{cr}} \\ \frac{1}{6} \log \xi(\chi), & \xi(\chi) \gtrsim \xi_{\mathrm{cr}} \end{cases} \tag{15}$$

Estimating the central charge using only a finite window of $\xi(\chi)$ one would then obtain an estimate qualitatively similar to the left panel of Fig. S11.

We see that such a qualitative picture is consistent with the overall behavior of the estimate of $c$ for $H^M$ (Fig. S11 middle panel). A similar discussion applies to $H^U$, where $c$ is supposed to jump from 1 to 0 as $J_\perp^U$ is increased. Also in this case, the smoothing of the jump at finite bond dimension is qualitatively consistent with the discussion above (Fig. S11 right panel).

Finally, as already stressed in the main text, we remark that for $H^U$ our simulations are not sufficient to reliably extract the central charge of the transition at $J_\perp^U \simeq 3.27$, which we nonetheless expect to be $c = 3/2$ from theoretical considerations (see main text).

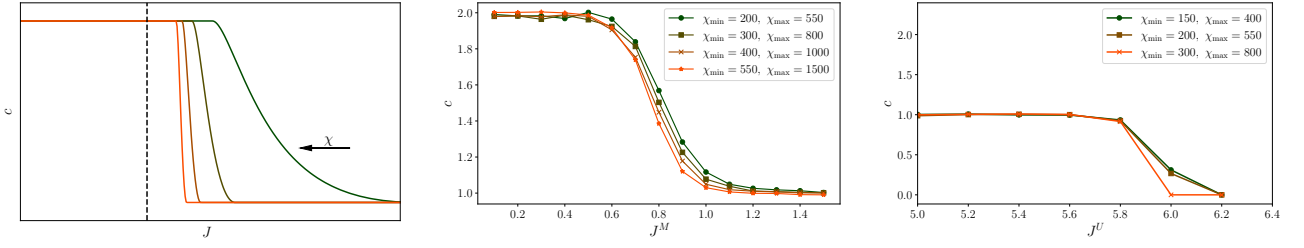

FIG. S11. (left panel) Sketch of the bond-dimension dependent smoothing of the jump in the estimated central charge, as explained in text. A dashed vertical line denotes $J_\perp^*$ where the $c$ would jump discontinuously in the $\chi \to \infty$ limit. (middle and right panels) Analysis of the convergence of the central charge $c$ on the bond dimension for $H^M$ (left) and $H^U$ (right). Each dot is obtained by using only a subset of all the available bond dimensions (see text), as specified by the range $[\chi_{\min}, \chi_{\max}]$.

## IV. MORE BOSONIZATION ANALYSIS

### A. Bosonization convention

Let us first begin by specifying the bosonization convention used throughout the work. This is most transparently done by stating the formulas for spin operators:

$$S^z_{\alpha j} \approx \frac{1}{2\pi}\partial_x \phi_\alpha + (-1)^j \sqrt{2}\mathcal{B}\sin\phi_\alpha + \dots \tag{16}$$

$$S^\pm_{\alpha j} \approx \exp\left(\pm i\theta_\alpha(x)\right)\left[\mathcal{C}(-1)^j + \mathcal{D}\cos\phi_\alpha(x)\right] + \dots \tag{17}$$

$\phi_\alpha \cong \phi_\alpha + 2\pi$ and $\theta_\alpha \cong \theta_\alpha + 2\pi$ are canonically conjugate pairs of compact bosons satisfying the commutation relation

$$[\partial_x \phi_\alpha(x), \theta_\alpha(y)] = 2\pi i \delta_{\alpha\beta}\delta(x-y). \tag{18}$$

The precise values of the real prefactors $\mathcal{B}, \mathcal{C}$ and $\mathcal{D}$ will not be important to us. Other details such as the handling of Klein factors follow standard treatments [12], but are not explicitly discussed here.

### B. Effective field theory for the critical surface of $H^M$

Let us begin by writing down the effective bosonized form of the Hamiltonian $H^M$.

$$H^M \approx \frac{v}{2\pi}\int dx \sum_{\alpha=1}^{2}\left[\frac{1}{4K}(\partial_x\phi_\alpha)^2 + K(\partial_x\theta_\alpha)^2\right] + \mathcal{A}^2\delta\int dx\, (\cos\phi_1 + \cos\phi_2) \\ + \mathcal{B}^2 J_\perp^M \int dx(\cos(\phi_1-\phi_2) - \cos(\phi_1+\phi_2)) + J_\perp^M\int dx\frac{\partial_x\phi_1 \partial_x\phi_2}{4\pi^2} \tag{19}$$

The action of various symmetries on the boson fields are as follows

1. $U(1)\times U(1)$ spin rotations: $\theta_\alpha \mapsto \theta_\alpha + \chi_\alpha,\ \phi_\alpha \mapsto \phi_\alpha$,
2. $\mathbb{Z}_2$ spin reflections: $\theta_\alpha \mapsto -\theta_\alpha,\ \phi_\alpha \mapsto -\phi_\alpha$,
3. $\mathbb{Z}_2$ Layer exchange: $\phi_1 \mapsto \phi_2,\ \theta_1 \mapsto \theta_2$.

As seen in the phase diagram shown in fig. S1, the critical surface of $H^M$, obtained by tuning $\delta = 0$, has three segments, two with central charge $c=1$ separated by a $c=2$ region. We use $\pm J_\perp^{M*}$ to indicate the values of $J_\perp^M$ when the critical surface of $H^M$ transitions from $c=2$ to $c=1$. A perturbative expression for scaling dimensions of $\cos(\phi_1 \pm \phi_2)$ can be written as

$$K_\pm \approx 2K\left(1 \mp \frac{J_\perp^M K}{2\pi v}\right). \tag{20}$$

This gives us $J_\perp^{M*} \approx (K-1)\frac{2\pi v}{K}$. When $-J_\perp^{M*} < J_\perp^M < J_\perp^{M*}$ all operators are irrelevant and the critical $c = 2$ theory can be thought of as a non-linear sigma model with the target space of a squashed torus

$$H \approx \frac{v}{2\pi} \int dx \sum_{\alpha=1}^{2} \left[ \frac{1}{4K} (\partial_x \phi_\alpha)^2 + K (\partial_x \theta_\alpha)^2 \right] + J_\perp^M \int dx \frac{\partial_x \phi_1 \partial_x \phi_2}{4\pi^2}. \quad (21)$$

When $J_\perp^M > J_\perp^{M*}$, $\cos(\phi_1 + \phi_2)$ becomes relevant and pins $\langle \phi_1 + \phi_2 \rangle = 0$. To describe the residual effective field theory, we use the following canonical $SL(2,\mathbb{Z})$ transformation

$$\varphi_1 = \phi_1 + \phi_2, \ \vartheta_1 = \theta_1, \ \varphi_2 = \phi_2, \ \vartheta_2 = \theta_2 - \theta_1 \quad (22)$$

which preserves both the compactification $\vartheta_\alpha \cong \vartheta_\alpha + 2\pi$, $\varphi_\alpha \cong \varphi_\alpha + 2\pi$ and the commutation in eq. (18),

$$[\partial_x \varphi_\alpha(x), \vartheta_\alpha(y)] = 2\pi i \delta_{\alpha\beta} \delta(x - y). \quad (23)$$

When $\varphi_1 = \phi_1 + \phi_2$ is pinned to $\langle \varphi_1 \rangle = 0$, we are left with an effective theory in terms of $\varphi_2$ and $\vartheta_2$ as follows

$$H \approx \frac{v_{\text{eff}}}{2\pi} \int dx \left[ \frac{1}{4K_{\text{eff}}} (\partial_x \varphi_2)^2 + K_{\text{eff}} (\partial_x \vartheta_2)^2 \right]. \quad (24)$$

The action of symmetries on the low-energy degrees of freedom are as follows

1. $U(1) \times U(1)$ spin rotations: $\vartheta_2 \mapsto \vartheta_2 + \chi_2 - \chi_1$, $\varphi_2 \mapsto \varphi_2$,
2. $\mathbb{Z}_2$ spin reflections: $\vartheta_2 \mapsto -\vartheta_2$, $\varphi_2 \mapsto -\varphi_2$,
3. $\mathbb{Z}_2$ Layer exchange: $\varphi_2 \mapsto -\varphi_2$, $\vartheta_2 \mapsto -\vartheta_2$.

On the other hand, when $J_\perp^M < -J_\perp^{M*}$, $\cos(\phi_1 - \phi_2)$ becomes relevant and pins $\langle \phi_1 - \phi_2 \rangle = 0$. To describe the residual effective field theory, we use a slightly different canonical $SL(2,\mathbb{Z})$ transformation

$$\varphi_1 = \phi_1 - \phi_2, \ \vartheta_1 = \theta_1, \ \varphi_2 = \phi_2, \ \vartheta_2 = \theta_1 + \theta_2 \quad (25)$$

When $\varphi_1 = \phi_1 - \phi_2$ is pinned to $\langle \varphi_1 \rangle = 0$, we are left with an effective theory in terms of $\varphi_2$ and $\vartheta_2$

$$H \approx \frac{v_{\text{eff}}}{2\pi} \int dx \left[ \frac{1}{4K_{\text{eff}}} (\partial_x \varphi_2)^2 + K_{\text{eff}} (\partial_x \vartheta_2)^2 \right]. \quad (26)$$

The action of symmetries on the low-energy degrees of freedom are as follows

1. $U(1) \times U(1)$ spin rotations: $\vartheta_2 \mapsto \vartheta_2 + \chi_1 + \chi_2$, $\varphi_2 \mapsto \varphi_2$,
2. $\mathbb{Z}_2$ spin reflections: $\vartheta_2 \mapsto -\vartheta_2$, $\varphi_2 \mapsto -\varphi_2$,
3. $\mathbb{Z}_2$ Layer exchange: $\varphi_2 \mapsto \varphi_2$, $\vartheta_2 \mapsto \vartheta_2$.

In summary, the three segments of the critical surface of $H^M$ are described by the theories shown in eqs. (21), (24) and (26). Observe that the two critical theories with $c = 1$ are inequivalent since they carry distinct symmetry charges (for example, $\varphi_2 \mapsto \mp \varphi_2$ and $\vartheta_2 \mapsto \mp \vartheta_2$ under layer exchange) and therefore cannot be connected which further justifies the presence of the intervening $c = 2$ region on the critical surface.

### C. Effective field theory for the critical surface of $H^U$

Let us begin by writing down the effective bosonized form of the Hamiltonian $H^U$

$$H^U \approx \frac{v}{2\pi} \int dx \sum_{\alpha=1}^{2} \left[ \frac{1}{4K} (\partial_x \phi_\alpha)^2 + K (\partial_x \theta_\alpha)^2 \right] + \mathcal{A}^2 \delta \int dx \ (\cos \phi_1 + \cos \phi_2) + \mathcal{C}^2 J_\perp^U \int dx \cos(\theta_1 - \theta_2)$$
$$+ \mathcal{B}^2 J_\perp^M \int dx \left( \cos(\phi_1 - \phi_2) - \cos(\phi_1 + \phi_2) \right) + J_\perp^M \int dx \frac{\partial_x \phi_1 \partial_x \phi_2}{4\pi^2} \quad (27)$$

The action of various symmetries on the boson fields are as follows

1. $U(1)$ spin rotations: $\theta_\alpha \mapsto \theta_\alpha + \chi$, $\phi_\alpha \mapsto \phi_\alpha$,

2. $\mathbb{Z}_2$ spin reflections: $\theta_\alpha \mapsto -\theta_\alpha,\ \phi_\alpha \mapsto -\phi_\alpha$,

3. $\mathbb{Z}_2$ Layer exchange: $\phi_1 \mapsto \phi_2,\ \theta_1 \mapsto \theta_2$.

As seen in the phase diagram fig. S4, the critical surface, obtained by tuning $\delta = 0$ consists of a $c = 1$ XY$_2$ line and two $c = 1$ XY$_1$ phases. We use $\pm J_\perp^{U*}$ to indicate the values of $J_\perp^U$ when the scaling dimensions $[\cos(\theta_1 - \theta_2)] = [\cos(\phi_1 - \phi_2)]$ and the critical surface of transitions from XY$_2$ to XY$_1$. Let us begin with $-J_\perp^{U*} < J_\perp^U < J_\perp^{U*}$ when $\cos(\phi_1 - \phi_2)$ pins $\langle \phi_1 - \phi_2 \rangle = 0$. This critical theory was already studied in section IV B and shown in eq. (26). The action of symmetries on the low-energy degrees of freedom are as follows

1. $U(1)$ spin rotations: $\vartheta_2 \mapsto \vartheta_2 + 2\chi,\ \varphi_2 \mapsto \varphi_2$,

2. $\mathbb{Z}_2$ spin reflections: $\vartheta_2 \mapsto -\vartheta_2,\ \varphi_2 \mapsto -\varphi_2$,

3. $\mathbb{Z}_2$ Layer exchange: $\varphi_2 \mapsto \varphi_2,\ \vartheta_2 \mapsto \vartheta_2$.

Now let us consider $|J_\perp^U| > J_\perp^{U*}$. $\cos(\theta_1 - \theta_2)$ becomes relevant and pins $\langle \theta_1 - \theta_2 \rangle$ to either 0 or $\pi$ for $J_\perp^U < 0$ and $J_\perp^U > 0$ respectively. The resulting effective theory can be obtained using the following canonical $SL(2,\mathbb{Z})$ transformation

$$\Theta_1 = \theta_1 - \theta_2,\ \Phi_1 = \phi_1,\ \Theta_2 = \theta_2,\ \Phi_2 = \phi_1 + \phi_2 \tag{28}$$

when $\Theta_1 = \theta_1 - \theta_2$ is pinned to any value, we are left with an effective theory in terms of $\Theta_2$ and $\Phi_2$

$$H \approx \frac{v_{\text{eff}}}{2\pi} \int dx \left[ \frac{1}{4K_{\text{eff}}} (\partial_x \Phi_2)^2 + K_{\text{eff}} (\partial_x \Theta_2)^2 \right]. \tag{29}$$

However, the action of symmetries on the low-energy fields depends on the value to which $\langle \Theta_1 \rangle$ is pinned to as follows

1. $U(1)$ spin rotations: $\Theta_2 \mapsto \Theta_2 + \chi,\ \Phi_2 \mapsto \Phi$,

2. $\mathbb{Z}_2$ spin reflections: $\Theta_2 \mapsto -\Theta_2,\ \Phi_2 \mapsto -\Phi_2$,

3. $\mathbb{Z}_2$ Layer exchange: $\Theta_2 \mapsto \Theta_2 + \langle \Theta_1 \rangle, \Phi_2 \mapsto \Phi_2$

We see that the local operator $e^{i\Theta}$ carries a trivial or non-trivial charge under layer exchange symmetry depending on whether $\langle \Phi_1 \rangle$ is pinned to 0 or $\pi$ and therefore represent distinct phases. In terms of the lattice operators, this is reflected in the nature of correlations of appropriate local operators. For example, $S_{1j}^+ + S_{2j}^+ \sim e^{i\Theta_2}\left(e^{i\langle\Theta_1\rangle} + 1\right)$ which carries trivial charge under layer exchange has two-point correlations that decay algebraically when $\langle \Theta_1 \rangle = 0$ and exponentially when $\langle \Theta_1 \rangle = \pi$. On the other hand, $S_{1j}^+ - S_{2j}^+ \sim e^{i\Theta_2}\left(e^{i\langle\Theta_1\rangle} - 1\right)$ which carries a non-trivial charge under layer exchange has two-point correlations that decay algebraically when $\langle \Theta_1 \rangle = \pi$ and exponentially when $\langle \Theta_1 \rangle = 0$.

### D. Topological response to gauge fields: the SPT phase in $H^M$

A useful diagnostic for SPT phases is studying their topological response to background gauge fields. This can be determined easily using the continuum bosonized theory. Consider $H^M$ whose continuum theory is shown in eq. (19). The phase diagram of $H^M$, shown in fig. S1 contains an SPT phase corresponding to when $\langle \phi_\alpha \rangle = \pi$. Since the system has two independent $U(1)$ symmetries, we can couple it to two independent $U(1)$ gauge fields to get a partition function written as a Euclidean path integral as follows [13]

$$\mathcal{Z}[A_1, A_2] = \int \mathcal{D}\phi_1\, \mathcal{D}\phi_2\, \exp[-S[\phi_1, \phi_2, A_1, A_2]] \tag{30}$$

$$S[\phi_1, \phi_2, A_1, A_2] = S[\phi_1, \phi_2, 0, 0] + \frac{i}{2\pi} \int_{\mathcal{M}_2} d^2x\, (\phi_1 F_1 + \phi_2 F_2) \tag{31}$$

where, $F_i \equiv \epsilon_{\mu\nu} \partial^\mu A_i^\nu$ is the dual electromagnetic field strength for the gauge field $A_i$. Let us assume that $\mathcal{M}_2$ has no boundaries. When $\langle \phi_\alpha \rangle = \pi$, we have

$$S[\phi_1, \phi_2, A_1, A_2] \approx S[\phi_1, \phi_2, 0, 0] + \frac{i}{2}\int_{\mathcal{M}_2} d^2x\, (F_1 + F_2) \approx S[\phi_1, \phi_2, 0, 0,] + i\pi\,(n_1 + n_2) \tag{32}$$

where we have used the flux quantization condition

$$\int_{\mathcal{M}_2} d^2x \ F_i = 2\pi n_i \tag{33}$$

and $n_\alpha \in \mathbb{Z}$ denotes the quanta of magnetic fluxes circulating in the closed Euclidean spacetime $\mathcal{M}_2$. This gives us the topological response of the SPT phase to background $U(1)$ gauge fields

$$\frac{\mathcal{Z}[A_1, A_2]}{\mathcal{Z}[0,0]} = (-1)^{n_1+n_2} . \tag{34}$$

We see that the response is non-trivial when $n_1 + n_2$ is odd. When we have open boundaries, $\partial \mathcal{M}_2 \neq 0$, we have

$$\mathcal{Z}[A_1, A_2] = \mathcal{Z}[0,0] \ \exp\left(\frac{i}{2}\int_{\partial\mathcal{M}_2} dx_\mu A_1^\mu + \frac{i}{2}\int_{\partial\mathcal{M}_2} dx_\mu A_2^\mu\right). \tag{35}$$

In other words, the response to each gauge field comes from the boundary via an anomalous 'improperly quantized' 0+1 dimensional Chern-Simons term for each gauge field. This signals the presence of stable boundary modes.

### E. Topological response to gauge fields: the region with boundary modes in $H^U$

We now consider $H^U$ which has a single gapped phase. However, depending on whether the physics is controlled by $XY_1$ or $XY_2$ which represent qualitatively distinct fixed points, the physics can be different. This will help us understand the stable boundary modes seen in the shaded region of fig. S4 above the $XY_2$ line. The system has a single $U(1)$ symmetry and thus we can couple it to a single $U(1)$ gauge field $A$ as follows

$$\mathcal{Z}[A] = \int \mathcal{D}\phi_1 \ \mathcal{D}\phi_2 \ \exp[-S[\phi_1, \phi_2, A]] \tag{36}$$

$$S[\phi_1, \phi_2, A] = S[\phi_1, \phi_2, 0] + \frac{i}{2\pi}\int_{\mathcal{M}_2} d^2x \ (\phi_1 + \phi_2) \ F \tag{37}$$

Let us begin outside the shaded region in fig. S4. This is described by $\langle \phi_1 + \phi_2 \rangle = 0$, $\langle \theta_1 - \theta_2 \rangle \neq 0$. The response to gauge fields becomes trivial:

$$S[\phi_1, \phi_2, A] \approx S[\phi_1, \phi_2, 0] + \frac{i}{2\pi}\int_{\mathcal{M}_2} d^2x \ \langle \phi_1 + \phi_2 \rangle \ F \approx S[\phi_1, \phi_2, 0] \tag{38}$$

In other words, the response is trivial indicating the absence of any boundary modes. On the other hand, inside the shaded region which is described by $\langle \phi_1 \rangle = \langle \phi_2 \rangle = \pi$, we have

$$S[\phi_1, \phi_2, A] \approx S[\phi_1, \phi_2, 0] + \frac{i}{2\pi}\int_{\mathcal{M}_2} d^2x \ \langle \phi_1 + \phi_2 \rangle \ F \approx S[\phi_1, \phi_2, 0] + i\int_{\mathcal{M}_2} d^2x \ F. \tag{39}$$

When $\mathcal{M}_2$ has no boundaries, we get

$$\int_{\mathcal{M}_2} d^2x \ F = 2\pi n \tag{40}$$

and therefore a trivial response:

$$\mathcal{Z}[A] = \mathcal{Z}[0] e^{2\pi i n} = \mathcal{Z}[0]. \tag{41}$$

With open boundary conditions, $\partial \mathcal{M}_2 \neq 0$, we get

$$\mathcal{Z}[A] = \mathcal{Z}[0] \exp\left(i\int_{\partial\mathcal{M}_2} dx^\mu A_\mu\right). \tag{42}$$

We see that the boundary responds via a properly quantized 0+1 dimensional Chern-Simons term. This is the response of a 0+1 dimensional SPT phase. Since $0+1$ dimensional SPT phases are classified by irreducible representations (irrep) of the symmetry group [14], this means that the boundary contains a non-trivial irrep of $O(2)$ since the $U(1)$ that is responding to the gauge field is contained in it. All non-trivial irreps of $O(2)$ are two-dimensional which shows that the edge modes are atleast two-fold degenerate on each end.

## F. String order parameters

We use bosonization to briefly explain the behaviour of various string order parameters used in this work. Let us begin with the string operators in eqs. (4) and (5) that can respectively distinguish between the trivial and non-trivial SPT phases of $H^M$. Their bosonized form can be written as [15, 16]

$$R^\alpha_{Triv} \sim \lim_{|x-y|\to\infty} \left\langle \cos\left(\frac{\phi_\alpha(x)}{2}\right) \cos\left(\frac{\phi_\alpha(y)}{2}\right) \right\rangle + \ldots, \tag{43}$$

$$R^\alpha_{SPT} \sim \lim_{|x-y|\to\infty} \left\langle \sin\left(\frac{\phi_\alpha(x)}{2}\right) \sin\left(\frac{\phi_\alpha(y)}{2}\right) \right\rangle + \ldots \tag{44}$$

We see that when $\langle \phi_\alpha \rangle = 0$, we have $R^\alpha_{Triv} \neq 0$ and $R^\alpha_{SPT} = 0$ whereas when $\langle \phi_\alpha \rangle = \pi$, we have $R^\alpha_{Triv} = 0$ and $R^\alpha_{SPT} \neq 0$ as seen in section I B.

The string order parameter used in the main text, $\mathcal{O}_{SC}$ that helped us distinguish between the different segments of the critical surface can be bosonized as [15]

$$\mathcal{O}_{SC} \sim \lim_{|x-y|\to\infty} A_+ \left\langle \cos\left(\frac{\phi_1(x)+\phi_2(x)}{2}\right) \cos\left(\frac{\phi_1(y)+\phi_2(y)}{2}\right) \right\rangle + A_- \left\langle \cos\left(\frac{\phi_1(x)-\phi_2(x)}{2}\right) \cos\left(\frac{\phi_1(y)-\phi_2(y)}{2}\right) \right\rangle + \ldots \tag{45}$$

We see that when either $\langle \phi_1 \pm \phi_2 \rangle = 0$, we get $\mathcal{O}_{SC} \neq 0$ as seen in the main text.

## V. PROXIMATE PHASE DIAGRAMS

We now look at phase diagrams of $H^M$ and $H^U$ proximate to the parameter regimes considered in the main text.

### A. Proximate phase diagrams of $H^M$

In the main text, we studied the phase diagram of $H^M$ eq. (1) with the value of $\Delta$ fixed to $\Delta \in \left(-\frac{1}{\sqrt{2}}, 0\right)$. We now sketch the phase diagrams for other values of $\Delta$. Recall that a useful way of determining our phase diagrams was to start with the decoupled limit $J^M_\perp \to 0$ where $H^M$ reduces to two independent copies of bond-dimerized XXZ chains and then studying the effect of the rung-couplings $J^M_\perp \neq 0$. We will use the same strategy here.

We begin by writing down the Hamiltonian of a single spin-$\frac{1}{2}$ XXZ chain with bond dimerization,

$$H_{XXZ,\delta} = \sum_{j=1}^{L} \left(1+(-1)^j\delta\right) \left[S^x_j S^x_{j+1} + S^y_j S^y_{j+1} + \Delta S^z_j S^z_{j+1}\right], \tag{46}$$

and sketching its phase diagram, which is shown in fig. S12. Qualitative features of this phase diagram can be found in [17] for $\delta > 0$ but can understood straightforwardly by bosonizing the Hamiltonian eq. (46) as

$$H \approx \frac{v}{2\pi} \int dx \left[\frac{1}{4K}\left(\partial_x\phi\right)^2 + K\left(\partial_x\theta\right)^2\right] + \mathcal{A}^2 \delta \int dx \, \cos\phi - \mathcal{B}^2 \Delta \int dx \, \cos 2\phi. \tag{47}$$

Here, the Luttinger parameter $K$ and velocity $v$ are obtained from the Bethe ansatz solution of the XXZ model [18] for $\Delta \in [-1, 1]$ as

$$K = \frac{\pi}{2\left(\pi - \arccos\Delta\right)}, \quad v = \frac{K}{2K-1}\sin\left(\frac{\pi}{2K}\right) \tag{48}$$

where $\mathcal{A}$ and $\mathcal{B}$ are the same unimportant coefficients that also appear in the main text. The various phases in fig. S12 can be understood by tracking the relevance of the operators $\cos\phi$ and $\cos 2\phi$ which have scaling dimensions $K$ and $4K$ respectively. In what follows, we will use fig. S12 to first determine the phase diagram for $J^{M/U}_\perp = 0$ line and then extrapolate the results to $J^{M/U}_\perp \neq 0$

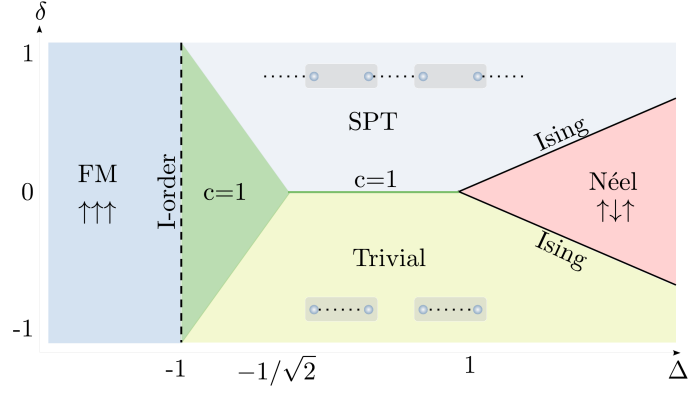

FIG. S12. Schematic phase diagram of a single XXZ spin chain with bond-dimerization, whose Hamiltonian is shown in eq. (46)

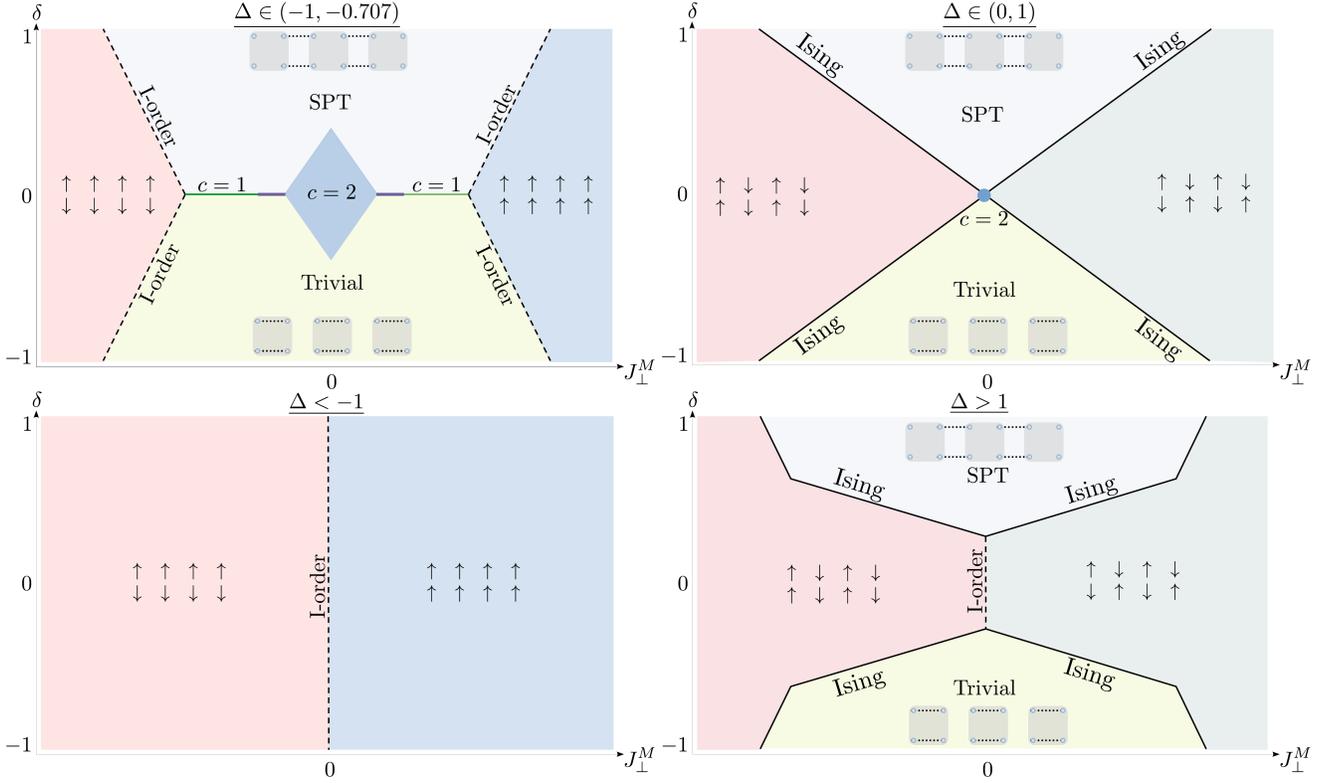

FIG. S13. Phase diagram of $H^M$ shown in eq. (49) for various values of $\Delta$.

Let us begin by writing down $H^M$ again,

$$H^M = \sum_{j=1}^{L}\sum_{\alpha=1}^{2} \left(1 + (-1)^j \delta\right) \left[S^x_{\alpha j} S^x_{\alpha j+1} + S^y_{\alpha j} S^y_{\alpha j+1} + \Delta S^z_{\alpha j} S^z_{\alpha j+1}\right] + J^M_\perp \sum_j S^z_{1j} S^z_{2j}, \qquad (49)$$

which can be bosonized as shown in eq. (19)

whose phase diagram we determined for $\Delta \in \left(-\frac{1}{\sqrt{2}}, 0\right)$ in the main text (also reproduced in the left panel of fig. S1). We now consider other values of $\Delta$. We will use the values of $K_\pm$ plotted in the right panel of fig. S1, which correspond to the scaling dimensions of $\cos(\phi_1 \pm \phi_2)$ whose perturbative form was shown in eq. (20). Another helpful piece of information is the nature of the ordered phase of $H^M$ (eq. (49)) in limit of strong interlayer coupling $J^M_\perp \to \pm\infty$. In

this limit, the dominant part of the Hamiltonian is

$$H^M \approx J_\perp^M \sum_j S_{1j}^z S_{2j}^z \tag{50}$$

whose ground state is extensively degenerate and is spanned by two states on each rung which we will label as $|\Uparrow\rangle \equiv |\uparrow_1\downarrow_2\rangle, |\Downarrow\rangle \equiv |\downarrow_1\uparrow_2\rangle$ for $J_\perp^M \to +\infty$ and $|\Uparrow\rangle \equiv |\uparrow_1\uparrow_2\rangle, |\Downarrow\rangle \equiv |\downarrow_1\downarrow_2\rangle$ for $J_\perp^M \to -\infty$. The nature of the phase for $J_\perp^M \to \pm\infty$ is determined by projecting the Hamiltonian $H^M$ in eq. (49) onto the ground space of eq. (50). This results in an effective Ising interaction to leading order.

$$H_{\text{eff}} \approx PHP = 2J_\perp^M \Delta \sum_j S_j^z S_{j+1}^z \text{ where } P = \prod_j \left(|\Uparrow_j\rangle\langle\Uparrow_j| + |\Downarrow_j\rangle\langle\Downarrow_j|\right). \tag{51}$$

The specific broken-symmetry ground states of $H_{\text{eff}}$ depend on the signs of $J_\perp^M$ and $\Delta$; for each choice, we have a two-fold degenerate ground state as listed below:

$$\Delta > 0 : |GS(J_\perp^M \to +\infty)\rangle = \left|\begin{array}{c} \cdots \uparrow \downarrow \uparrow \downarrow \cdots \\ \cdots \downarrow \uparrow \downarrow \uparrow \cdots \end{array}\right\rangle, \left|\begin{array}{c} \cdots \downarrow \uparrow \downarrow \uparrow \cdots \\ \cdots \uparrow \downarrow \uparrow \downarrow \cdots \end{array}\right\rangle, \tag{52}$$

$$\Delta > 0 : |GS(J_\perp^M \to -\infty)\rangle = \left|\begin{array}{c} \cdots \uparrow \downarrow \uparrow \downarrow \cdots \\ \cdots \uparrow \downarrow \uparrow \downarrow \cdots \end{array}\right\rangle, \left|\begin{array}{c} \cdots \downarrow \uparrow \downarrow \uparrow \cdots \\ \cdots \downarrow \uparrow \downarrow \uparrow \cdots \end{array}\right\rangle, \tag{53}$$

$$\Delta < 0 : |GS(J_\perp^M \to +\infty)\rangle = \left|\begin{array}{c} \cdots \uparrow \uparrow \uparrow \uparrow \cdots \\ \cdots \downarrow \downarrow \downarrow \downarrow \cdots \end{array}\right\rangle, \left|\begin{array}{c} \cdots \downarrow \downarrow \downarrow \downarrow \cdots \\ \cdots \uparrow \uparrow \uparrow \uparrow \cdots \end{array}\right\rangle, \tag{54}$$

$$\Delta < 0 : |GS(J_\perp^M \to -\infty)\rangle = \left|\begin{array}{c} \cdots \uparrow \uparrow \uparrow \uparrow \cdots \\ \cdots \uparrow \uparrow \uparrow \uparrow \cdots \end{array}\right\rangle, \left|\begin{array}{c} \cdots \downarrow \downarrow \downarrow \downarrow \cdots \\ \cdots \downarrow \downarrow \downarrow \downarrow \cdots \end{array}\right\rangle. \tag{55}$$

Using these pieces of information, we can determine the proximate phase diagrams for various $\Delta$:

- $\Delta \in (0, 1)$: From fig. S12, we see that when $J_\perp^M \to 0$, we have the trivial and non-trivial SPT phases for $\delta \neq 0$ separated by a $c = 2$ transition. From eqs. (20) and (48), we see that since $K \in \left(\frac{1}{2}, 1\right)$, we have $K_\pm \in (1, 2)$ and therefore both operators $\cos(\phi_1 \pm \phi_2)$ are relevant for $J_\perp^M \neq 0$. This drives the system to the phases obtained for large $|J_\perp^M|$ which have ground states shown in eqs. (52) and (53). Since both ordered ground states belong to the same $S_{tot,1}^z = \sum_j S_{1j}^z$ and $S_{tot,2}^z = \sum_j S_{2j}^z$ sectors as the trivial and non-trivial SPT phases, the transition between the latter phases and the ordered phases is expected to be second-order and belong to the Ising universality class [19]. Putting all these together, we get the phase diagram shown in fig. S13 (top, right).

- $\Delta \in \left(-1, -\frac{1}{\sqrt{2}}\right)$: From fig. S12, we see that when $J_\perp^M \to 0$, we have the trivial and non-trivial SPT phases for $\delta \to \pm 1$ separated by a $c = 2$ line around $\delta = 0$ where $\cos\phi_{1,2}$ are irrelevant. For $J_\perp^M \neq 0$, no relevant operators are introduced and the $c = 2$ critical line extends to a $c = 2$ critical phase. For $\delta \neq 0$, this phase terminates when $\cos\phi_{1,2}$ become relevant and drives the system to the trivial or non-trivial SPT phase depending on the sign of $\delta$. For $\delta = 0$, the $c = 2$ line persists until one of $\cos(\phi_1 \pm \phi_2)$ becomes relevant, depending on the sign of $J_\perp^M$ and changes the gapless phase to $c = 1$ which describes a transition between the trivial and non-trivial SPT phases. For large $|J_\perp^M|$, we get the ordered phases with ground states shown in eqs. (54) and (55). Since both ordered ground states belong to different $S_{tot,1}^z = \sum_j S_{1j}^z$ and $S_{tot,2}^z = \sum_j S_{2j}^z$ sectors as compared to the trivial and non-trivial SPT phases, the transition between the latter phases and the ordered phases is expected to be first-order. Putting all these together, we get the phase diagram shown in fig. S13 (top, left).

- $\Delta > 1$: From fig. S12, we see that when $J_\perp^M \to 0$, we have the trivial and non-trivial SPT phases for $\delta \to \pm 1$ separated by an ordered system with four degenerate ground states shown in eqs. (52) and (53). This represents a first-order phase transition between the ordered phases with two-fold degenerate ground states eq. (52) for $J_\perp^M > 0$ and eq. (53) for $J_\perp^M < 0$. The ordered phases are straddled by the trivial and non-trivial SPT phases with a second-order transition between them belonging to the Ising universality class [19]. The phase diagram is shown in fig. S13 (bottom, right).

- $\Delta < -1$: From fig. S12, we see that when $J_\perp^M \to 0$, we have an ordered system with four degenerate ground states shown in eqs. (54) and (55) for all $\delta$. This represents a first-order phase transition between the ordered phases with two-fold degenerate ground states eq. (54) for $J_\perp^M > 0$ and eq. (55) for $J_\perp^M < 0$. The phase diagram is shown in fig. S13 (bottom, left).

### B. Proximate phase diagrams of $H^U$

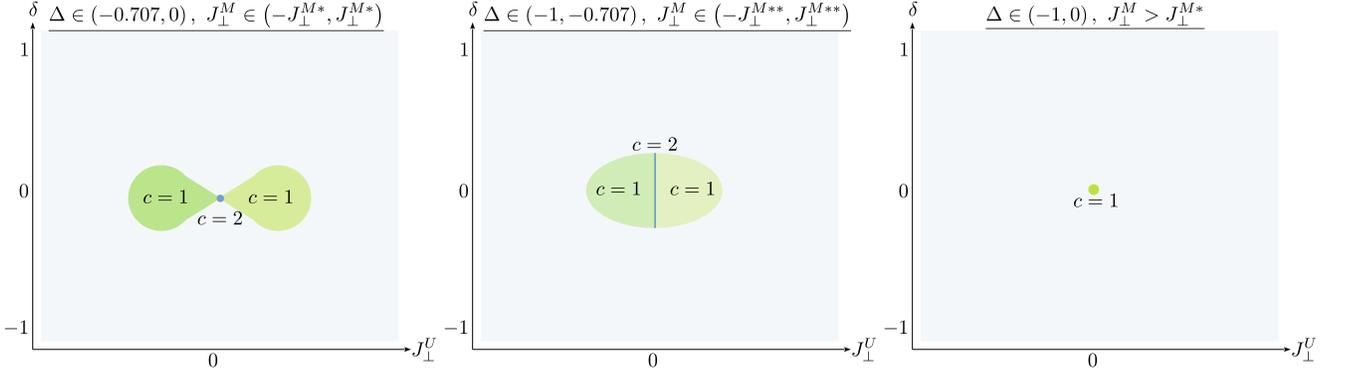

FIG. S14. Phase diagram of $H^U$ shown in eq. (56) for various values of $\Delta$ and $J_\perp^M$.

We now move on to $H^U$ whose form we write below

$$H^U = \sum_{j=1}^{L} \sum_{\alpha=1}^{2} \left(1 + (-1)^j \delta\right) \left[S_{\alpha j}^x S_{\alpha j+1}^x + S_{\alpha j}^y S_{\alpha j+1}^y + \Delta S_{\alpha j}^z S_{\alpha j+1}^z\right] + J_\perp^U \sum_j \left(S_{1j}^x S_{2j}^x + S_{1j}^y S_{2j}^y\right) + J_\perp^M S_{1j}^z S_{2j}^z, \quad (56)$$

which can be bosonized as shown in eq. (27), whose phase diagram we determined for $\Delta \in (-1, 0)$ and $J_\perp^M < J_\perp^{M*}$ in the main text (also shown in the left panel of fig. S4). We now consider phase diagrams for other values of $\Delta \in (-1, 1)$ and $J_\perp^M$ (restricting ourselves to the range of parameters such that there exists a critical point at $\delta = 0$). We use $\pm J_\perp^{M*}$ to indicate the values of $J_\perp^M$ beyond which the $\delta = 0$ theory for $H^M$ transitions from $c = 2$ to $c = 1$.

First, let us note that the edges of the phase diagram can be determined to consist of a single trivial phase as in the main text. The differences between the phase diagrams all occur in the region surrounding the origin. The $J_\perp^U = 0$ line corresponds to a cut across the phase diagram of $H^M$ shown in fig. S13 with a fixed value of $J_\perp^M$. This helps us determine the phase diagrams of $H^U$ by extrapolation.

- $\Delta \in \left(-\frac{1}{\sqrt{2}}, 0\right)$, $J_\perp^M \in \left(-J_\perp^{M*}, J_\perp^{M*}\right)$: for $J_\perp^U = 0$, the phase transition at $\delta = 0$ corresponds to a $c = 2$ point where both $\cos(\phi_1 \pm \phi_2)$ are irrelevant. For $J_\perp^U \neq 0$, the relevant operator $\cos(\theta_1 - \theta_2)$ reduces the theory to an XY$_1$ phase with $c = 1$. Consider the dimerization operator ,

$$\cos \phi_1 + \cos \phi_2 = \cos\left(\frac{(\phi_1 + \phi_2) + (\phi_1 - \phi_2)}{2}\right) + \cos\left(\frac{(\phi_1 + \phi_2) - (\phi_1 - \phi_2)}{2}\right). \quad (57)$$

Since $\cos(\theta_1 - \theta_2)$ pins $\theta_1 - \theta_2$, $\phi_1 - \phi_2$ fluctuates and a two-point correlation function of $\cos\phi_1 + \cos\phi_2$ decays exponentially. Therefore, dimerization is not a scaling operator and no longer opens up a gap. As a result, just as in the phase diagram of the main text, the XY$_1$ theory opens up into lobes of a critical phase as shown in the leftmost panel of fig. S14.

- $\Delta \in \left(-1, -\frac{1}{\sqrt{2}}\right)$, $J_\perp^M \in \left(-J_\perp^{M**}, J_\perp^{M**}\right)$: $\pm J_\perp^{M**}$ denotes the point on the $\delta = 0$ line where the $c = 2$ phase changes to a line in the top, left figure of fig. S13. The $J_\perp^U = 0$ line corresponds to a cut across the $c = 2$ phase in the phase diagram shown in the upper left panel of fig. S13. Thus, the $c = 2$ point at the origin of the phase diagram for $\Delta \in \left(-\frac{1}{\sqrt{2}}, 0\right)$, $J_\perp^M \in \left(-J_\perp^{M*}, J_\perp^{M*}\right)$ opens up into a line as shown in the middle panel of fig. S14.

- $\Delta \in (-1, 0)$, $J_\perp^M > J_\perp^{M*}$ (note that the case studied in the main text is instead $J_\perp^M < -J_\perp^{M*}$): for $J_\perp^U = 0$, the phase transition at $\delta = 0$ corresponds to a $c = 1$ point where $\cos(\phi_1 + \phi_2)$ is relevant while $\cos(\phi_1 - \phi_2)$ is irrelevant. For $J_\perp^U \neq 0$, $\cos(\theta_1 - \theta_2)$ is relevant and immediately gaps the theory out to the trivial phase. This gives us a phase diagram with an isolated $c = 1$ unnecessary critical point at the origin as shown in the rightmost panel of fig. S14.

We end by observing that the two lobes of XY$_1$ are always separated by an intermediate $c = 1$ XY$_2$ regime or a $c = 2$ phase transition consistent with the analysis of section IV which shows that they are distinct phases.

### C. Other possible terminations of the unnecessary critical line

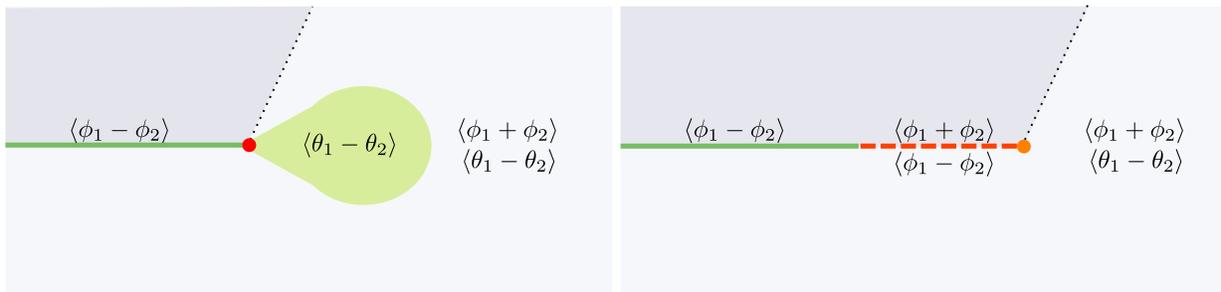

FIG. S15. Two scenarios by which the unnecessary critical line can terminate. Fields in angular brackets are pinned. The thick continuous and broken lines represent second and first order phase transitions respectively. Shaded regions denote the presence of boundary modes and the thin dotted lines denotes a boundary transition. Left: Scenario 1 which was realized in the phase diagram of $H^U$. The solid line and extended bulb denotes an $XY_2$ line and $XY_1$ phase respectively both with $c=1$. The dot where the boundary transition terminates corresponds to a $c=\frac{3}{2}$ critical theory corresponding to an Ising CFT stacked on top of a compact boson. Right: Scenario 2 where the critical $XY_2$ line changes to a first-order line via a KT transition and then terminates via an Ising transition (orange dot). The boundary transition is expected to meet the bulk critical line at the Ising point.

Finally, let us now look at other ways the unnecessary critical line could have terminated. Recall that the trivial phase corresponds to $\theta_1 - \theta_2$ and $\phi_1 + \phi_2$ being pinned. In the phase diagram shown in the left panel of fig. S4, the $XY_2$ critical line, where only $\phi_1 - \phi_2$ was pinned first transitioned to an $XY_1$ critical phase where $\theta_1 - \theta_2$ was pinned until the scaling dimension $K_+$ reduced sufficiently that $\cos(\phi_1 + \phi_2)$ became relevant pinning $\phi_1 + \phi_2$. Let us call this scenario 1. A phase diagram realizing this is schematically shown in the left panel of fig. S15. An alternative possibility is that $\cos(\phi_1 + \phi_2)$ first becomes relevant, pinning $\phi_1 + \phi_2$ along with $\phi_1 - \phi_2$. This changes the second-order unnecessary critical line to first-order, which can then terminate when $\theta_1 - \theta_2$ gets pinned instead of $\phi_1 - \phi_2$ through an Ising transition [19]. Let us call this scenario 2. A schematic phase diagram of this scenario is shown in the right panel of fig. S15. While it has not been realized in the parameter regimes of the model we studied, a scenario similar to this was seen in the unnecessary critical phase diagram in Ref.[7]. Stable boundary modes are expected to exist above the critical line so long as $\phi_1 - \phi_2$ is pinned. It is curious that in both scenarios, the boundary transition terminates on the unnecessary critical line, at an Ising transition.

## VI. THE 'FAILED SPT' PREMISE

The phase diagrams for multiversality and unnecessary criticality were constructed using a template built from the results of the classification of symmetry protected topological (SPT) phases [14, 20]. In this section, we briefly review important results from this program and how they were used in producing the phase diagrams studied in the main text. To keep the discussion concrete, we will restrict our attention to SPT phases of 1+1 D bosons with unitary on-site symmetries.

### A. The classification of 1+1 D bosonic SPT phases

To begin, let us recall that a 1+1 D Hamiltonian $H$ which has a global symmetry $G$ is said to belong to a non-trivial SPT phase protected by $G$ if $H$ has a unique gapped ground state with periodic boundary and degeneracies arising from protected boundary modes with open boundary conditions. Furthermore, we require that the non-trivial properties of $H$ are parametrically stable to any weak deformations $H \mapsto H + \varepsilon V$ using any symmetry preserving perturbations $V$. For example, the Haldane phase of the Hamiltonian in eq. (46) shown in fig. S12 can be thought of as a non-trivial SPT phase protected by an on-site $O(2)$ symmetry discussed in the main text. The Haldane phase is said to have a $\mathbb{Z}_2$ classification. This means two things: (i) A stack of two copies of a Hamiltonian belonging to the Haldane phase can be adiabatically connected to the trivial SPT phase, in a path preserving a diagonal $O(2)$ symmetry without encountering a phase transition. This adiabatic path was explicitly demonstrated with $H^U$, the model for unnecessary criticality (eq. (6)) (ii) The Haldane phase is the only non-trivial 1+1 D bosonic SPT phase protected by an on-site unitary $O(2)$ symmetry.

More generally, 1+1 dimensional bosonic SPT phases protected by a unitary on-site symmetry $G$ are completely classified by $\mathcal{H}^2(G, U(1))$, the so-called second group cohomology group [14, 20]. $\mathcal{H}^2(G, U(1))$ is an Abelian group whose elements label distinct SPT phases protected by $G$ where, in particular, the identity element labels the trivial phase. The group structure of $\mathcal{H}^2(G, U(1))$ encodes the information about the 'stacking rules' of SPT phases i.e. given two Hamiltonians belonging to distinct SPT phases, what SPT phase the composite system obtained by stacking the two belongs to. The Haldane phase and trivial phase in fig. S12 corresponds to the trivial and non-trivial group elements of $\mathcal{H}^2(O(2), U(1)) \cong \mathbb{Z}_2$.

## B. Unwinding 1+1 D bosonic SPT phases

A powerful device to understand aspects of SPT phases is to enumerate the conditions under which they become trivial, which we will call unwinding. Below, we enumerate various ways in which this can be achieved. More details can be found in Refs.[21–23]

*Inversion*: This is essentially related to the group structure of the classification of SPT phases described above. Given a Hamiltonian belonging to a non-trivial SPT phase, labelled by some non-trivial element of $[\omega] \in \mathcal{H}^2(G, U(1))$, by stacking another Hamiltonian belonging to the inverse element $[\omega^{-1}] \in \mathcal{H}^2(G, U(1))$, the composite system can be deformed to a trivial phase without breaking the diagonal $G$ symmetry.

*Symmetry breaking*: By explicitly breaking the $G$ symmetry down to one of its subgroups $H \subset G$, a Hamiltonian belonging to a non-trivial $G$ SPT phase can be adiabatically connected to the trivial phase. For example, by breaking the spin reflection symmetry of the Hamiltonian eq. (46) by adding a magnetic field $h \sum_j S_j^z$, the $O(2)$ symmetry is reduced to $U(1)$ and the Haldane phase can be connected to the trivial one.

*Symmetry extension*: By appropriately extending the $G$ symmetry to an appropriately larger symmetry $\tilde{G}$ (eg: by embedding the bosonic Hilbert space into a fermionic one), a Hamiltonian belonging to a non-trivial $G$ SPT phase can be adiabatically connected to the trivial phase. For example, by extending the $O(2)$ symmetry of the Haldane phase of eq. (46) to its double cover[24], $Pin(2)$, it can be connected to the trivial phase.

Let us now present a slightly more mathematical exposition of the above results. A group homomorphism $\varphi$ between two groups $G_1$ and $G_2$

$$\varphi : G_1 \to G_2 \tag{58}$$

induces a group homomorphism $\varphi^*$ between $\mathcal{H}^2(G_2, U(1))$ and $\mathcal{H}^2(G_1, U(1))$ by pullback as follows

$$\varphi^* : \mathcal{H}^2(G_2, U(1)) \to \mathcal{H}^2(G_1, U(1)). \tag{59}$$

Concretely, for $\{g_1, h_1\} \in G_1$, we have $\{\varphi(g_1), \varphi(h_1)\} \in G_2$. Given a set of representative cocycles [14] $\omega(g_2, h_2) \in \mathcal{H}^2(G_2, U(1))$, we can get $\varphi^* \omega(g_1, h_1) = \omega(\varphi(g_1), \varphi(h_1))$ where $\varphi^* \omega(g_1, h_1) \in \mathcal{H}^2(G_1, U(1))$.

Consider the case when $H$ is a subgroup of $G$. This means that there exists an injective group homomorphism

$$i : H \to G, \tag{60}$$

which induces a group homomorphism between $\mathcal{H}^2(G, U(1))$ and $\mathcal{H}^2(H, U(1))$

$$i^* : \mathcal{H}^2(G, U(1)) \to \mathcal{H}^2(H, U(1)). \tag{61}$$

A $G$-SPT phase can be unwound by explicitly breaking symmetry to $H$ if the image of the group element of $\mathcal{H}^2(G, U(1))$ the SPT phase belongs to is the trivial element in eq. (61). This is always possible, for example by choosing to break all symmetries i.e. $H \cong \{1\}$.

Now, consider the case when there exists a surjective group homomorphism between $\tilde{G}$ and $G$

$$s : \tilde{G} \to G, \tag{62}$$

which induces a group homomorphism between $\mathcal{H}^2(G, U(1))$ and $\mathcal{H}^2(\tilde{G}, U(1))$

$$s^* : \mathcal{H}^2(G, U(1)) \to \mathcal{H}^2(\tilde{G}, U(1)). \tag{63}$$

A $G$-SPT phase can be unwound by extending the symmetry to $\tilde{G}$ if the image of the group element of $\mathcal{H}^2(G, U(1))$ the SPT phase belongs to, is the trivial element in eq. (63). While the condition is relatively easy to phrase, the proof that there always exists atleast one suitable $\tilde{G}$ is non-trivial (see Refs. [21, 23, 25, 26]).

Finally, consider two non-trivial $G$-SPT phases labelled by some non-trivial elements $\omega_1, \omega_2 \in \mathcal{H}^2(G, U(1))$. A decoupled stack of Hamiltonians belonging to the two phases has a $G \times G$ symmetry whose SPT phases are classified by $\mathcal{H}^2(G \times G, U(1))$, which can be expanded using the Künneth formula as

$$\mathcal{H}^2(G \times G, U(1)) \cong \mathcal{H}^2(G, U(1)) \times \mathcal{H}^2(G, U(1)) \times \mathcal{H}^1(G, \mathcal{H}^1(H, U(1))). \tag{64}$$

The composite system belongs to a non-trivial $G \times G$ SPT phase labelled by the element $\{\omega_1, \omega_2, 1\}$ for each group in the expansion on the right hand side of eq. (64). Now consider the diagonal subgroup $G \in G \times G$ which gives us the following injective group homomorphism

$$i : G \to G \times G \tag{65}$$

and furnishes the following pullback group homomorphism

$$i^* : \mathcal{H}^2(G \times G, U(1)) \to \mathcal{H}^2(G, U(1)). \tag{66}$$

To see which SPT phase the composite system belongs to when the $G \times G$ symmetry is broken to a diagonal $G$, we need to determine the image of $\{\omega_1, \omega_2, 1\} \in \mathcal{H}^2(G \times G, U(1))$ under $i^*$ in eq. (66). This is easily determined to be

$$i^*\left(\{\omega_1, \omega_2, 1\}\right) = \omega_1 \omega_2. \tag{67}$$

Equation (67) essentially recovers the stacking rule for SPT phases. In particular, when $\omega_2 = \omega_1^{-1}$, we get

$$i^*\left(\{\omega_1, \omega_1^{-1}, 1\}\right) = 1, \tag{68}$$

which tells us how to understand unwinding SPT phases via inversion.

### C. From unwinding SPT phases to unnecessary criticality

In the main text, the Hamiltonian in eq. (6) can be understood starting in the totally decoupled limit $J_\perp^{M,U} = 0$ where the model reduces to two copies of eq. (46). From the discussions above in section VI B, since $J_\perp^{M,U} \neq 0$ breaks the $O(2) \times O(2)$ down to a diagonal $O(2)$ and thus, the $J_\perp^{M,U} = 0$, $\delta > 0$ region which has boundary modes is a 'failed SPT' and can be unwound (via inversion) to connect to the $\delta < 0$ region without any bulk phase transitions. Under the right conditions as described in the main text, we can stabilize an unnecessary critical region on the $\delta = 0$ surface. A failed SPT premise based on unwinding by inversion has been used to produce higher dimensional field theoretic models hosting a phase diagram with unnecessary criticality in Refs. [27, 28].

The other ways of unwinding SPT phases can also serve as a failed-SPT template to build phase diagrams with unnecessary criticality. For example, the possibility of unwinding the Haldane phase via symmetry extension, by embedding the spins into a fermionic Hilbert space was originally noted by Anfuso and Rosch [29] and more recently explored in Ref.[7, 30]. All these phase diagrams contain unnecessary criticality.

### D. Resurrected SPT phases and multiversality

In the main text, we considered eq. (1) which can be understood starting in the decoupled limit $J_\perp^M \to 0$ where the model reduces to two copies of eq. (46) which has $O(2) \times O(2)$ symmetry. The distinction between the $\delta < 0$ and $\delta > 0$ gapped phases however is retained for $J_\perp^M \neq 0$ which preserves a $(U(1) \times U(1)) \rtimes \mathbb{Z}_2$ subgroup of $O(2) \times O(2)$. The edge modes are reduced in number on each end and the system crosses over from a $O(2) \times O(2)$ SPT phase to a $(U(1) \times U(1)) \rtimes \mathbb{Z}_2$ SPT phase. The Hamiltonian eq. (1) is obtained in a submanifold of the Hamiltonian space of eq. (6) where the failed SPT phase of eq. (6) is resurrected. Under the right conditions discussed in the main text, we see that the critical surface exhibits multiversality.

On a more technical level, we can understand this within the group cohomology framework in section VI B. We have the following injective group homomorphism

$$i : (U(1) \times U(1)) \rtimes \mathbb{Z}_2 \to O(2) \times O(2), \tag{69}$$

which induces the following group homomorphism via pullback

$$i^* : \mathcal{H}^2(O(2) \times O(2), U(1)) \to \mathcal{H}^2((U(1) \times U(1)) \rtimes \mathbb{Z}_2, U(1)). \tag{70}$$

The fact that breaking symmetry from $O(2) \times O(2)$ to $(U(1) \times U(1)) \rtimes \mathbb{Z}_2$ leaves the system behind in a non-trivial SPT phase means that the image of the non-trivial group element of $\mathcal{H}^2(O(2) \times O(2), U(1))$ representing the $\delta > 0, J_\perp^M = 0$ SPT phase under the map $i^*$ is a non-trivial element of $\mathcal{H}^2((U(1) \times U(1)) \rtimes \mathbb{Z}_2, U(1))$.

## VII. GENERALIZATION TO $2N$-LEG LADDERS

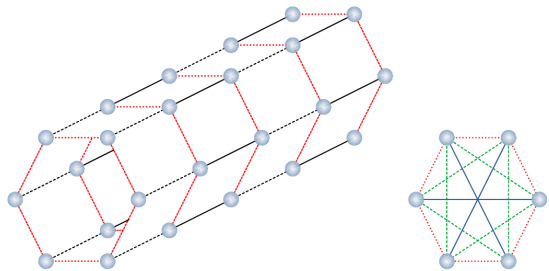

FIG. S16. Schematic representation of the Hamiltonians in eqs. (71) and (72) for N=3. The various lines denote Hamiltonian couplings.

The two-leg ladder models studied so far can easily be generalized to $2N$ legs arranged in the form of a symmetric tube as shown in fig. S16. The Hamiltonians take the form

$$H_N^M = \sum_{j=1}^{L} \sum_{\alpha=1}^{2N} (1 + \delta(-1)^j) \left[ S_{\alpha j}^x S_{\alpha j+1}^x + S_{\alpha j}^y S_{\alpha j+1}^y + \Delta S_{\alpha j}^z S_{\alpha j+1}^z \right] + \sum_{j=1}^{L} \sum_{k=1}^{N} J_{\perp,k}^M \sum_{\alpha=1}^{2N} S_{\alpha j}^z S_{\beta j}^z, \tag{71}$$

$$H_N^U = H_N^M + \sum_{j=1}^{L} \sum_{k=1}^{N} J_{\perp,k}^U \sum_{\substack{\alpha=1 \\ \beta=\alpha+k}}^{2N} \left[ S_{\alpha j}^x S_{\beta j}^x + S_{\alpha j}^y S_{\beta j}^y \right]. \tag{72}$$

where $\alpha \in 1, \ldots, 2N$ labels the legs of the ladder, with $\alpha + 2N \cong \alpha$ identified. A generalization of the bosonization arguments used for the two-leg models [31, 32] suggests that $H_N^M$ and $H_N^U$ exhibit multiversality and unnecessary criticality in appropriate parameter regimes. Since the symmetries preserved are now larger, the critical surface would be richer.

Let us begin with the totally decoupled limit $J_{\perp,k}^{M,U} = 0$. The Hamiltonian is now symmetric under independent spin rotations and reflections on each leg generating an on-site $O(2)^{2N}$ symmetry group, while the $\mathbb{Z}_2$ layer exchange symmetry is now enhanced to $S_{2N}$ permutations of the legs. Restricting $\Delta$ to $\Delta \in \left(-\frac{1}{\sqrt{2}}, 0\right)$, yields similar results in the decoupled limit to the two-leg case i.e. a trivial gapped phase for $\delta < 0$ and a collection of Haldane SPT paramagnets for $\delta > 0$, but now separated at $\delta = 0$ by a $c = 2N$ conformal field theory (CFT) corresponding to a $2N$-component Luttinger liquid. Switching on $J_{\perp,k}^M \neq 0$ reduces the $S_{2N}$ layer permutation symmetry to $\mathbb{Z}_{2N}$ and on-site symmetries to $U(1)^{2N} \rtimes \mathbb{Z}_2$. Bosonization [31, 32] tells us that the $c = 2N$ theory is perturbatively stable for small $|J_{\perp,k}^M|$. In appropriate parameter regimes, depending on the relevance of the vertex operators introduced by $J_{\perp,k}^M$, we expect a critical surface with multiversality driven by a cascade of transitions to conformal field theories with $c < 2N$.

On the other hand, also switching on $J_{\perp,k}^U \neq 0$ preserves the $\mathbb{Z}_{2N}$ layer exchange symmetry but reduces the on-site symmetry to a diagonal $O(2)$ generated by simultaneous spin rotations and reflections. This eliminates the distinction between the $\delta < 0$ and $\delta > 0$ gapped phases of the $J_{\perp,k}^U = 0$ limit. The phase diagram on the periphery $J_{\perp,k}^U \to \infty$ and $|\delta| = 1$ can be determined to comprise of a single adiabatically connected trivial phase just as before. We expect the region near the origin, $J_{\perp,k}^U \approx 0$, $\delta \approx 0$ to exhibit unnecessary criticality in appropriate parameter regimes, including stable boundary modes. We also expect the boundary transition to merge with the unnecessary critical surface.

## VIII. A POSSIBLE SYNTHESIS OF MORE GENERAL PHASE DIAGRAMS WITH MULTIVERSALITY

In the examples studied in the main text, we considered multiversality on the critical surface separating a trivial from a non-trivial SPT phase. An interesting possibility to consider is whether multiversality is possible on a critical surface separating other, possibly more conventional phases, such as those characterized by spontaneous symmetry breaking. In Ref.[33], the authors study field theoretic models where the critical theory of the spontaneous breaking of an Ising symmetry did not correspond to the expected Ising universality class, but a more complex deconfined quantum critical [34] theory— a phenomenon they dub 'Landau-beyond-Landau'. An interesting possibility that we could envision is a single critical surface corresponding to an Ising symmetry breaking phase transition with multiple

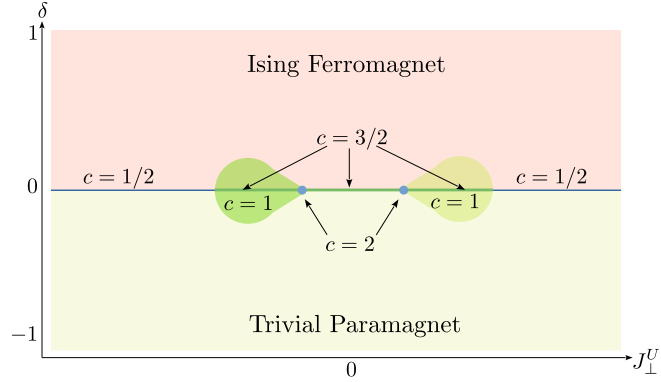

FIG. S17. Phase diagram of the Hamiltonian shown in eq. (73). The $c = \frac{1}{2}$ segments on the $\delta = 0$ critical surface represents the Ising universality class whereas $c = \frac{3}{2}$ represents a critical theory containing both Ising CFT as well as a compact boson.

universality classes, one of which is the usual Ising universality class. Here, we present a fine-tuned version of such phase diagrams. These are obtained by stacking a model with unnecessary criticality containing a trivial phase, on top of *any* phase diagram in such a way that the critical surfaces coincide. Since the model with unnecessary criticality does not change the phases of the model it is stacked upon, the only effect it will have is on the critical surface. For example, consider a three-leg spin $\frac{1}{2}$ ladder with the following Hamiltonian

$$H = \sum_{j=1}^{L} \sum_{\alpha=1}^{2} \left(1 + (-1)^j \delta\right) \left[S^x_{\alpha j} S^x_{\alpha j+1} + S^y_{\alpha j} S^y_{\alpha j+1} + \Delta S^z_{\alpha j} S^z_{\alpha j+1}\right] + J^U_\perp \sum_j \left(S^x_{1j} S^x_{2j} + S^y_{1j} S^y_{2j}\right) + J^M_\perp \sum_j S^z_{1j} S^z_{2j}$$
$$- \sum_{j=1}^{L} \left((1 - \delta) S^x_{3j} + 2(1 + \delta) S^z_{3j} S^z_{3j+1}\right). \quad (73)$$

This is nothing but the transverse field Ising model stacked on top of the Hamiltonian with unnecessary criticality, $H^U$ defined in eq. (6). The phase diagram of this model is shown in fig. S17 for fixed values of $\Delta$ and $J^M_\perp$ that gives us unnecessary criticality for $H^U$. We see that the critical surface separating the Ising paramagnet and ferromagnet contains, in addition to the Ising universality class, a 'beyond-Landau' critical theory with central charge $c = \frac{3}{2}$ corresponding to the compact boson CFT stacked on the Ising CFT.

At this stage, the model eq. (73) and its phase diagram in fig. S17 is fine-tuned and its stability to perturbations is unclear. An intriguing possibility is that by imposing suitable symmetries, aspects of the phase diagram fig. S17 coming from eq. (73) or a similar model can be made stable to generic symmetry-preserving perturbations. In this case, it is no longer a fine-tuned phenomenon and would constitute an example of a 'Landau-beyond-Landau' multiversal phase diagram.


	\fi

\end{document}